\documentclass{aa}  
\usepackage{graphicx}
\usepackage{txfonts}
\usepackage[tight,footnotesize]{subfigure}
\usepackage[center]{caption}
\usepackage{multirow}
\usepackage{xcolor}
\usepackage{comment}
\usepackage{natbib}
\usepackage[flushleft]{threeparttable}
\usepackage{placeins}
\usepackage{hyperref}
\hypersetup{
    colorlinks=true,
    linkcolor=blue,
    filecolor=magenta,      
    urlcolor=blue,
    citecolor=blue,
}

\begin{document} 
\title{Indications of magnetic accretion in Swift J0826.2-7033\thanks{Based on observations obtained with XMM-Newton, an ESA science mission with instruments and contributions directly funded by ESA Member States and NASA.}}
\author{ Nikita Rawat
          \inst{1},
          Domitilla De Martino 
          \inst{2},
          Koji Mukai
          \inst{3,4},
          Maurizio Falanga
          \inst{5,6},
          Nicola Masetti
          \inst{7,8},
          and Jeewan C. Pandey
          \inst{1}
}

\institute{Astronomy Division, Aryabhatta Research Institute of observational sciencES (ARIES), Nainital 263001, India\\
\email{rawatnikita221@gmail.com, nikita@aries.res.in}
\and 
Istituto
Nazionale di Astrofisica, Osservatorio Astronomico di Capodimonte, Salita Moiariello 16, I-80131, Napoli, Italy \\
\email{domitilla.demartino@inaf.it}
\and
CRESST II and X-ray Astrophysics Laboratory, NASA/GSFC, Greenbelt, MD 20771, USA
\and
Department of Physics, University of Maryland, Baltimore County, 1000 Hilltop Circle, Baltimore, MD 21250, USA
 \and 
 International Space Science Institute (ISSI), Hallerstrasse 6, CH-3012, Bern, Switzerland
 \and
 Physikalisches Institut, University of Bern, Sidlerstrasse 5, 3012 Bern, Switzerland
 \and
Istituto Nazionale di Astrofisica, Osservatorio di Astrofisica e Scienza dello Spazio di Bologna, Via Gobetti 101, I-40129, Bologna, Italy 
\and
Instituto de Astrof\'isica, Facultad de Ciencias Exactas, Universidad Andr\'es Bello, Fern\'andez Concha 700, Las Condes, Santiago, Chile
        }



\newcommand{\orb}{${\Omega}$\,} 
\newcommand{\twoorb}{${2\Omega}$\,} 
\newcommand{\threeorb}{${3\Omega}$\,} 
\newcommand{\po}{P$_{\Omega}$\,} 
\newcommand{\ps}{P$_{\omega}$\,} 
\newcommand{\pb}{P$_{\omega-\Omega} $\,} 
\newcommand{\ptwoo}{P$_{2\Omega}$\,} 
\newcommand{\pthreeo}{P$_{3\Omega}$\,} 
\newcommand{\pone}{P$_{1}$}
\newcommand{\pthree}{P$_{3}$}
\newcommand{\rn}[1]{%
  \textup{\uppercase\expandafter{\romannumeral#1}}%
}
 
\abstract
{We present our findings from the first long X-ray observation of the hard X-ray source Swift J0826.2-7033 with XMM-Newton, which has shown characteristics of magnetic accretion. The system appears to have a long
orbital period ($\sim$7.8 h) accompanied by short timescale variabilities, which we tentatively interpret as the spin and beat periods of an intermediate polar. These short- and long-timescale modulations are energy-independent, suggesting that photoelectric absorption does not play any role in producing the variabilities. If our suspected spin and beat periods are true, then Swift J0826.2-7033 accretes via disc-overflow with an equal fraction of accretion taking place via disc and stream. The XMM-Newton and Swift-BAT spectral analysis reveals that the post-shock region in Swift J0826.2-7033 has a multi-temperature structure with a maximum temperature of $\sim$43 keV, which is absorbed by a material with an average equivalent hydrogen column density of $\sim$1.6$\times10^{22}$ cm$^{-2}$ that partially covers $\sim$27\% of the X-ray source. The suprasolar abundances, with hints of an evolved donor, collectively make Swift J0826.2-7033 an interesting target, which likely underwent a thermal timescale mass transfer phase.
}

\keywords{accretion, accretion discs – novae, cataclysmic variables – X-rays: stars – stars: individual: Swift J0826.2-7033, 1RXS J082623.5-703142, PBC J0826.3-7033}

\titlerunning{Indications of magnetic accretion in Swift J0826.2-7033}
\authorrunning{N. Rawat et al.}
\maketitle

\section{Introduction}
Intermediate polars (IPs) represent a subclass of magnetic cataclysmic variables (MCVs) where the low magnetic field strength (B $\sim$1-10 MG) white dwarf (WD) accretes material from a Roche lobe filling secondary star \citep[see][for a full review of IPs]{1983ASSL..101..155W, 1994PASP..106..209P, 1995ASPC...85..185H}. These are asynchronous binaries, which means that two prominent periods, the orbital period of the binary system (\po) and the spin period of WD (\ps), with \ps $<$ \po, are characteristic features of this subclass. The weak magnetic field strength of the WD may allow the formation of a truncated accretion disc. Consequently, accretion on to the WD can occur either via a disc or a stream, or a combination of both. Thus, depending on the magnetic field strength of WD, mass accretion rate, and binary orbital separation, three accretion scenarios are thought to occur in IPs: disc-fed \citep{1988MNRAS.231..549R}, stream-fed \citep{1986MNRAS.218..695H}, and disc-overflow \citep{1995ASPC...85..185H}. In the disc-fed accretion, the accretion flow from the secondary proceeds towards the WD via a truncated accretion disc. While in the stream-fed accretion, an accretion disc does not form, and the material accretes directly from an accretion stream. The disc-overflow is a combination of disc-fed and stream-fed, where both accretion scenarios simultaneously exist. In MCVs, the accreting material is channelled along the magnetic field lines to the WD surface and as it reaches supersonic velocities, it forms a shock above the WD surface, which is hot ($\sim$10-50 keV), below which the matter cools down via bremsstrahlung radiation emitting in the hard X-rays and optical/near-IR cyclotron radiation \citep{1973PThPh..49.1184A}. The hard X-rays are reflected by either the WD surface or reprocessed by cool material above the pre-shock. Also, the X-ray and cyclotron emissions are partially thermalized by the surface of the WD, which emits as a blackbody in the soft X-rays and/or EUV/UV domains with relative proportion depending on the magnetic field strength \citep{1996A&A...306..232W}.

\par Swift J0826.2-7033 (hereafter J0826) was identified as an X-ray source in Swift-BAT hard X-ray sky survey \citep{2010A&A...510A..48C}. It was proposed as a likely non-magnetic CV based on optical spectral features by \cite{2012A&A...545A.101P}. They also estimated a WD mass of 0.4 M$_{\odot}$. J0846 is located at a distance of 366.3 $\pm$ 2.0 pc \citep{2021AJ....161..147B}. The G-band magnitude, as measured by Gaia, is 14.1. It has been a decade since its discovery, yet our understanding of this system is limited. In this paper, we present the findings from pointed XMM-Newton observations of J0826.  Our aim was to assess the variability characteristics in both X-ray and optical/UV domains, along with the X-ray spectral characteristics of this relatively unexplored CV.

\section{XMM-Newton observations and data reduction}
On 2018 April 24 at 21:30:24 (UT) at an offset of 1.94 arcmin, the XMM-Newton satellite \citep{2001A&A...365L...1J} observed J0826 (observation ID: 0820330401) with the European Photon Imaging Camera \citep[EPIC;][]{2001A&A...365L..18S, 2001A&A...365L..27T}  operated in small window mode with the thin filter. The EPIC has three detectors: two Metal Oxide Semiconductors (MOS1 and MOS2) and one p-n detector (PN).
The total exposure duration for the EPIC-PN and EPIC-MOS were 41.1 ks and 41.6 ks, respectively. J0826 was also observed with the Reflection Grating Spectrographs (RGS1 and RGS2; \citealt{2001A&A...365L...7D}) with a total exposure time of 41.9 ks.  We have utilized the standard XMM-Newton \textsc{science analysis system (sas)} software package (version 20.0.0) with the most recent calibration files available at the time of analysis\footnote{\url{https://www.cosmos.esa.int/web/xmm-newton/current-calibration-files}} while adhering to the \textsc{sas} analysis thread\footnote{\url{https://www.cosmos.esa.int/web/xmm-newton/sas-threads}} for data reduction.  We corrected event arrival times to the barycenter of the Solar System and inspected the data against high background flaring activity and found it not to be affected. We also looked for the occurrence of pile-up and found no evidence of it. To extract the final light curves and spectra from the event files, we selected a circular region with a 30$^{\prime\prime}$ radius centred on the source and a circular background region with the same size as that of the source. In this way, the light curves and spectra were extracted in the energy range of 0.2-12.0 keV and 0.3-10.0 keV, respectively. The average net count rates in MOS and PN light curves are 6.01 counts s$^{-1}$ and 10.42 counts s$^{-1}$, respectively. 

\par  In the case of RGS, only the first order (0.35-2.5 keV) spectra were utilized for RGS1 and RGS2. The average net count rate for RGS is 0.189 counts s$^{-1}$. Furthermore, temporal and spectral analyses were performed with \textsc{heasoft} version 6.29. The EPIC and RGS spectra have been rebinned to have 20 counts per bin in order to apply $\chi^{2}$ statistics.  J0826 was also observed using the optical monitor \citep[OM;][]{2001A&A...365L..36M} for a total exposure time of 35.6 ks in imaging fast mode in the B filter (3900-4900 $\AA$). A total of 10 different exposures were barycentric corrected and then combined to create the final summed light curve file. The average net count rates for OM data are 45.40 $\pm$ 0.04 counts s$^{-1}$, equivalent to an instrumental magnitude of 15.123 $\pm$ 0.001  mag\footnote{The zero-point value of 19.2661 for the B filter was used from the XMM-Newton Users Handbook available at \url{https://xmm-tools.cosmos.esa.int/external/xmm_user_support/documentation/uhb/XMM_UHB.pdf}. }. 

\begin{figure}[htb!]
\centering
\includegraphics[width=9cm, height=5.5cm]{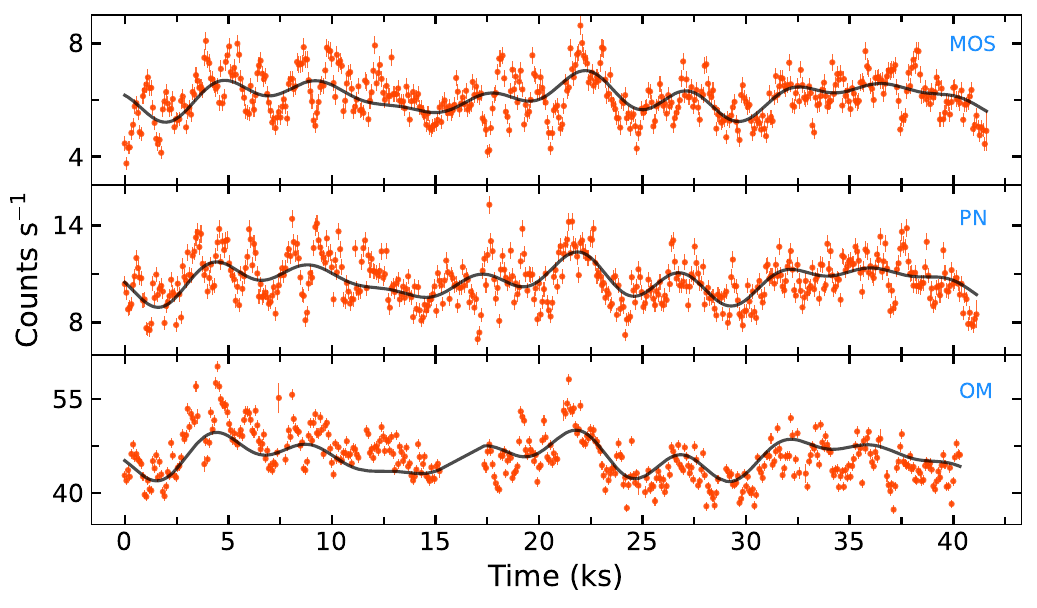}
\caption{EPIC light curves in the 0.2–12.0 keV energy range and OM light curve in B filter binned in 80 s intervals for clarity purposes. The black solid line represents the multi-fit sinusoidal function comprising three frequencies present in the X-ray light curves. } 
\label{fig:lc_xmm}
\end{figure}

\begin{figure*}[h!]
\centering
\subfigure[]{\includegraphics[width=8cm, height=6cm]{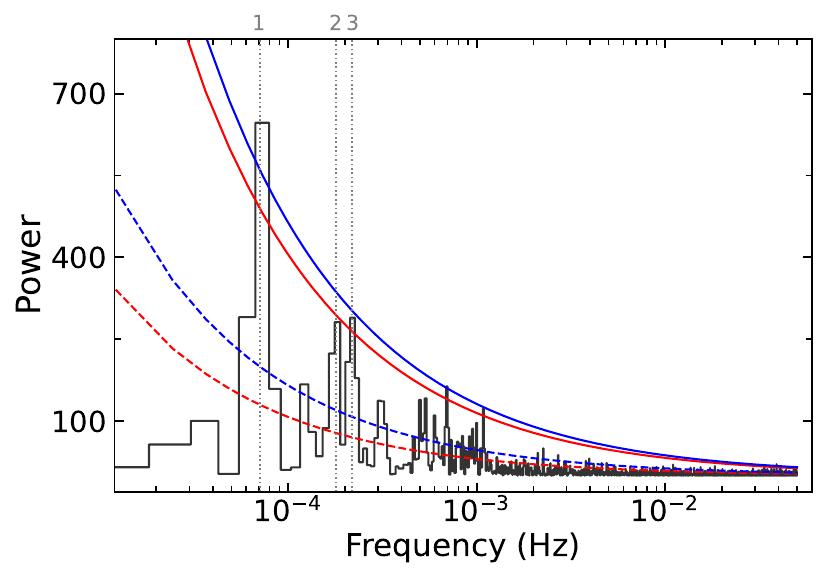} \label{fig:pn_ps}}
\subfigure[]{\includegraphics[width=8cm, height=6cm]{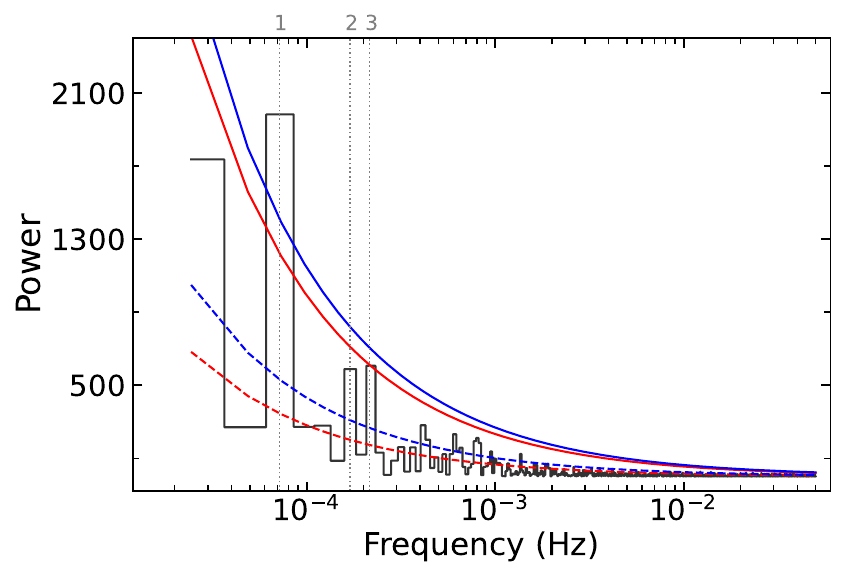} 
\label{fig:om_ps}}
\caption{Power spectra of J0826 obtained from (a) PN and (b) OM data. The red dashed and solid lines represent 95\% single trial and global significance levels, respectively, while the blue dashed and solid lines represent 99\% single trial and global significance levels, respectively. }
\label{fig:ps_xmm}
\end{figure*}

\section{Timing analysis}
\subsection{Light curves and power spectra}
The background-subtracted combined MOS (MOS1 and MOS2), PN, and OM light curves show short-term and long-term variability (see Fig. \ref{fig:lc_xmm}). The power spectra of light curves with a binning of 10 s were computed and we show the PN and OM power spectra in Fig. \ref{fig:ps_xmm}. From the power spectral analysis, three dominant peaks in the PN data are found (1, 2, and 3 denoted with grey vertical lines in Fig.  \ref{fig:ps_xmm} ), corresponding to periods P$_{1}$ = 14103 $\pm$ 149 s, P$_{2}$ = 5604 $\pm$ 34 s, and P$_{3}$ = 4588 $\pm$ 25 s. The OM data also reveal three peaks corresponding to periods P$_{1}$ = 13909 $\pm$ 120 s, P$_{2}$ = 5577 $\pm$ 31 s, and P$_{3}$ = 4633 $\pm$ 23 s.  The black solid line in Fig. \ref{fig:lc_xmm} represents the multi-fit sinusoidal function comprising of three frequencies present in the light curves. The uncertainties in these periods were determined by performing 10000 iterations of Markov Chain Monte Carlo (MCMC) simulations within the Period04 package \citep{2004IAUS..224..786L}.  In order to check for the significance levels, we have followed the method given by \cite{2005A&A...431..391V}.  The red dashed and solid lines in Fig. \ref{fig:ps_xmm} represent 95\% single trial and global significance levels, respectively, while the blue dashed and solid lines represent 99\% single trial and global significance levels, respectively. It is clear from Fig. \ref{fig:ps_xmm} that all three periods in the PN data are found to be significant at the 95\% global significance level. We also found two more peaks at slightly above the 95\% global significance level in the power spectrum of PN data, corresponding to periods $\sim$1434 s and $\sim$922 s. However, these periods are found in a cluster of an excess of power that precludes a detailed analysis. 

\subsection{Folded light curves} \label{sec3.1}
\begin{figure*}[htb!]
\centering
\subfigure[]{\includegraphics[width=8cm, height=9cm]{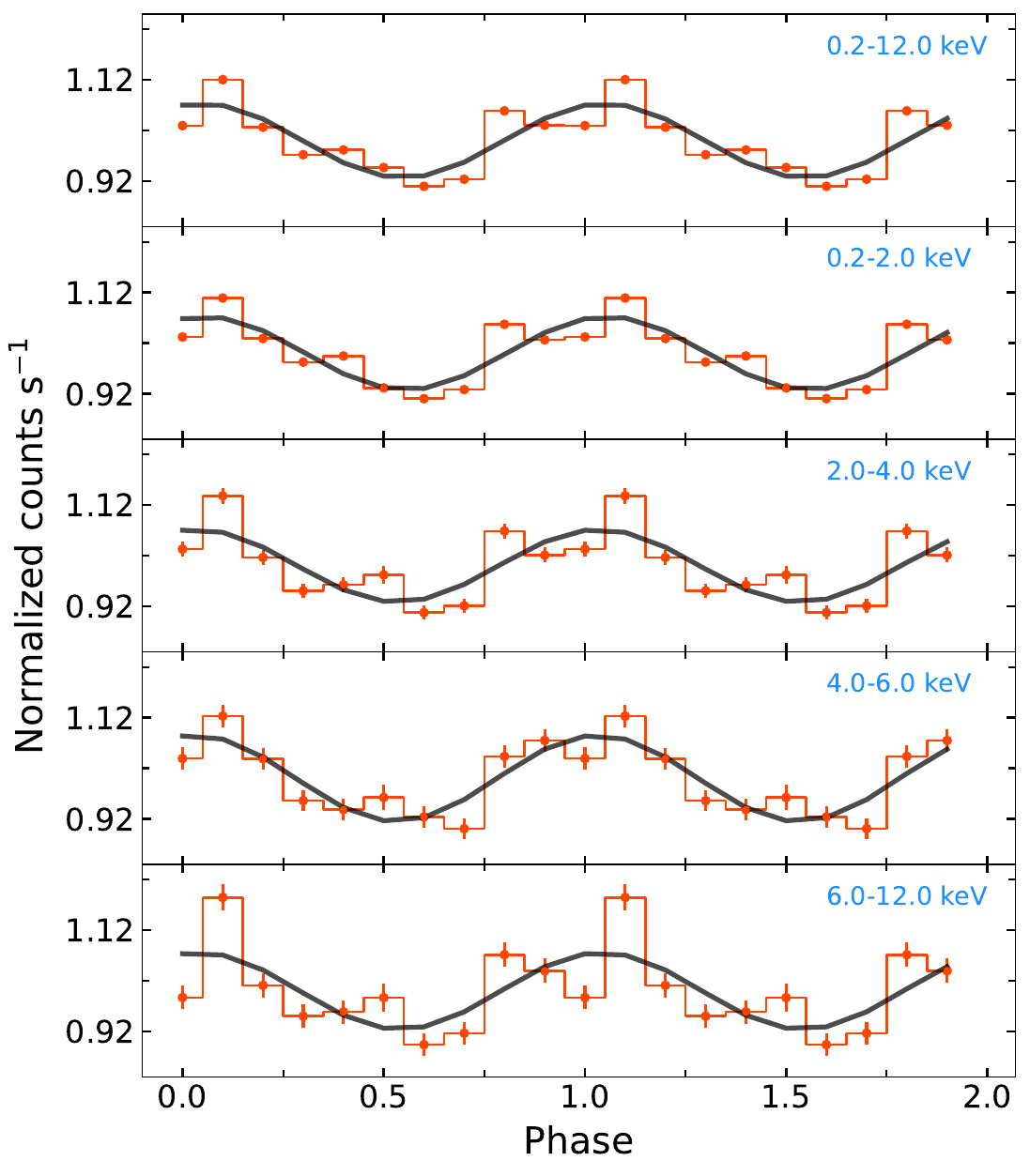} \label{fig:pn_ed_p1}}
\subfigure[]{\includegraphics[width=8cm, height=9cm]{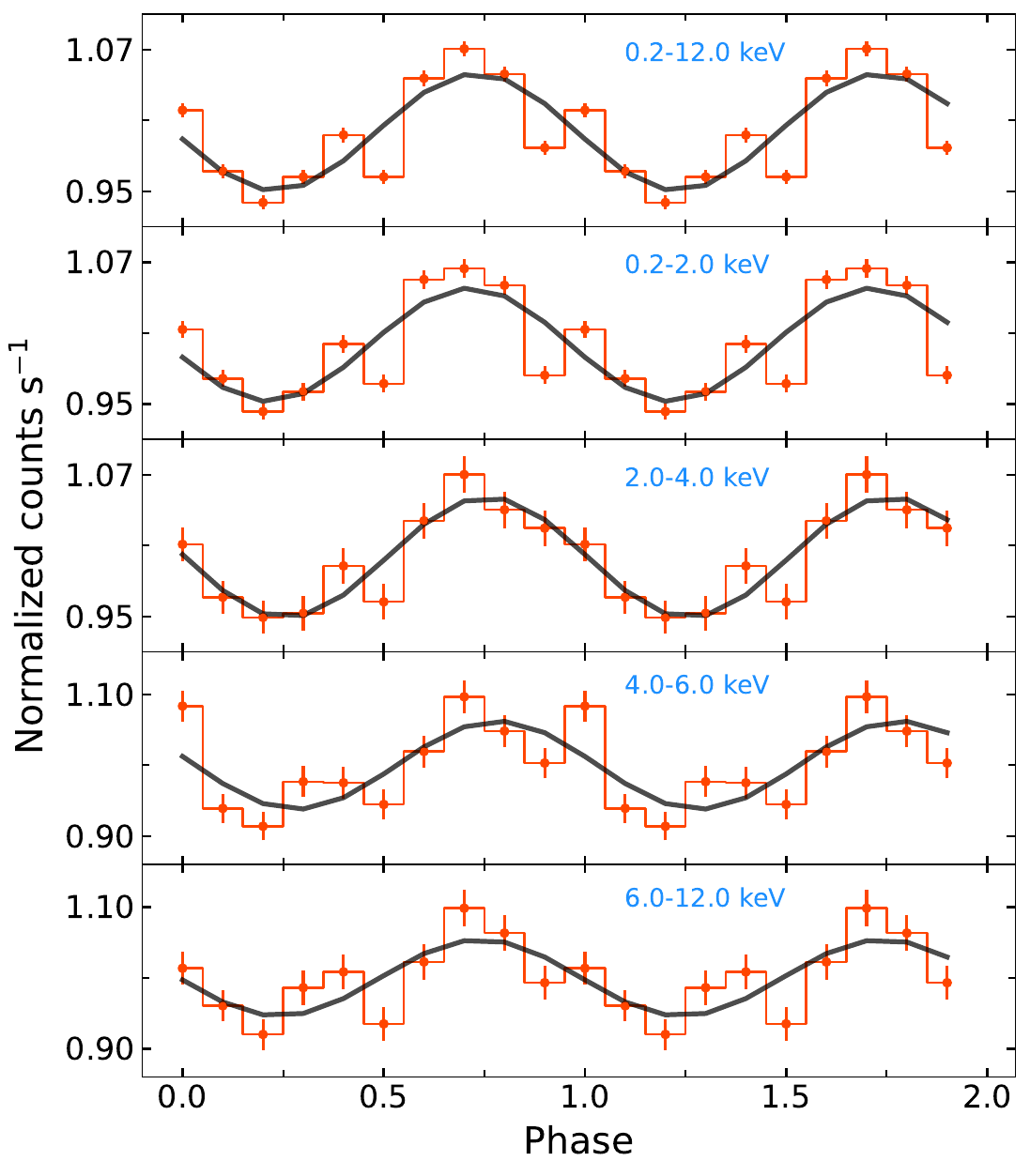} 
\label{fig:pn_ed_p2}}
\subfigure[]{\includegraphics[width=8cm, height=9cm]{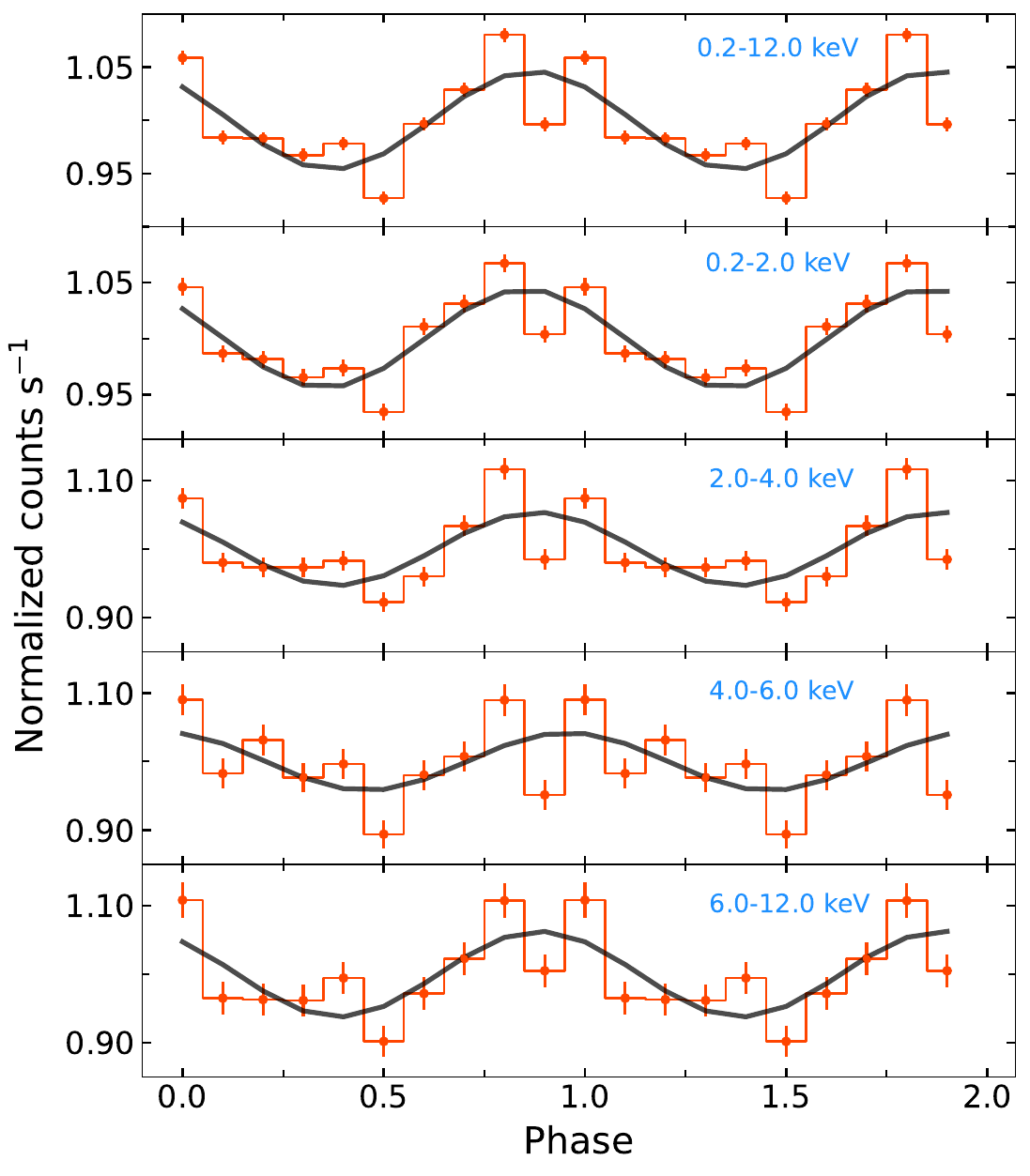}
\label{fig:pn_ed_p3}}
\subfigure[]{\includegraphics[width=8cm, height=9cm]{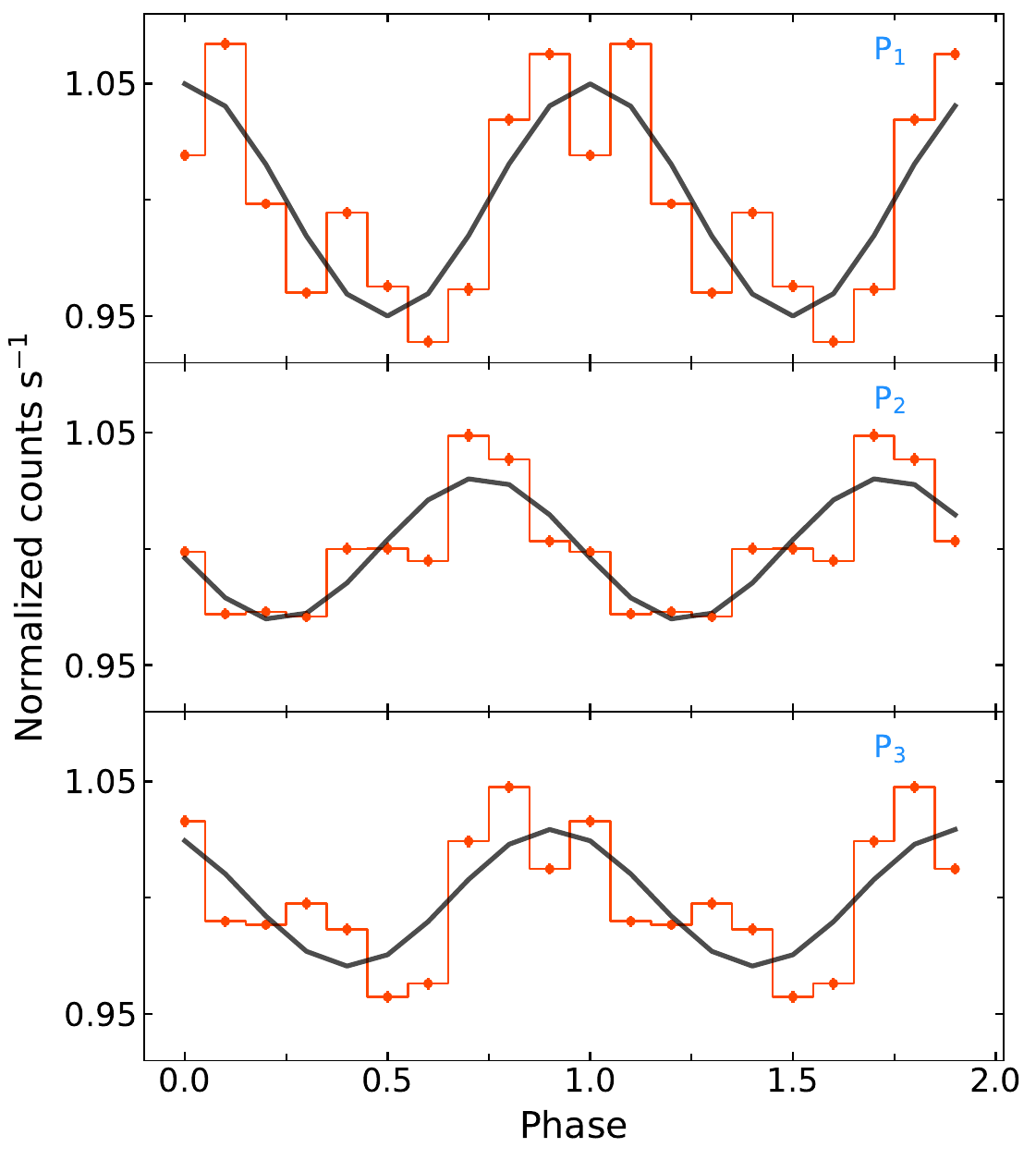} 
\label{fig:om_folded}}
\caption{ Energy-dependent folded light curves of PN data at periods (a) P$_{1}$ (b) P$_{2}$, (c) \pthree, and (d) OM folded light curves at periods P$_{1}$, P$_{2}$, and P$_{3}$. }
\label{fig:pn_om_folded}
\end{figure*}

We have folded energy-resolved PN light curves (0.2-2.0 keV, 2.0-4.0 keV, 4.0-6.0, and 6.0-12.0 keV) at periods P$_{1}$, P$_{2}$, and P$_{3}$ to check for the energy dependency (see, Figs. \ref{fig:pn_ed_p1}-\ref{fig:pn_ed_p3}). The reference time for folding was taken as MJD 58232, corresponding to the observation date of 2018-04-24 at 0 hr UT. The light curves were folded with a binning of 10 points in a phase. We have also derived the fractional amplitude for \pone, P$_{2}$, and \pthree ~by fitting a sinusoidal function to these light curves, in which the amplitude of the sinusoidal function denotes the derived value of fractional amplitudes. The derived values are given in Table \ref{tab:pulse-fraction}, which are well within a 1$\sigma$ level with each other. As can be seen from the folded light curves (Figs. \ref{fig:pn_ed_p1} and \ref{fig:pn_ed_p3}) and Table \ref{tab:pulse-fraction}, no explicit energy dependency is present, which indicates that the photoelectric absorption in the accretion flow, which is typical in IPs, might not be playing a role for these variabilities. Inspection of the hardness ratio curves at periods \pone ~and \pthree ~indeed does not show any variability. Moreover, the folded light curves at periods $\sim$1434 s and $\sim$922 s do not seem to show any significant variability.

\par The B-band light curve was also folded at the X-ray periods \pone, P$_{2}$, and \pthree ~with a binning of 10 points in a phase, which are shown in Fig. \ref{fig:om_folded}. The derived values of fractional amplitudes are 5.0 $\pm$ 0.8\%, 3.1 $\pm$ 0.4\%, and 2.9 $\pm$ 0.6\% for \pone, P$_{2}$, and \pthree, respectively. The optical light curves are broadly consistent with the phasing of the X-ray light curves for periods \pone ~and \pthree ~with maxima and minima of optical pulses centred on the X-ray maxima and minima, suggesting a common region where they are formed.

\begin{table}[h]
\centering
\caption{Energy-dependent fractional amplitude for periods P$_{1}$, P$_{2}$,  and P$_{3}$ obtained from the folded light curves.}
\label{tab:pulse-fraction}
\renewcommand{\arraystretch}{1.4}
\begin{tabular}{ccccccc}
\hline
\multirow{2}{*}{Energy Bands} && \multicolumn{5}{c}{Amplitude (\%)} \\
&& P$_{1}$  &&  P$_{2}$ && P$_{3}$ \\
\hline
0.2-12.0 && 7.4 $\pm$ 1.2 && 5.0 $\pm$ 0.8 && 4.6 $\pm$ 0.9\\
0.2-2.0 && 7.3 $\pm$ 1.1 && 4.8 $\pm$ 0.8 && 4.4 $\pm$ 0.7\\
2.0-4.0 && 7.2 $\pm$ 1.5 && 5.1 $\pm$ 0.6 && 5.4 $\pm$ 1.4 \\
4.0-6.0 && 8.5 $\pm$ 1.2 && 6.2 $\pm$ 1.3 && 4.2 $\pm$ 1.6  \\
6.0-12.0 && 7.6 $\pm$ 2.0 && 5.4 $\pm$ 1.2 && 6.2 $\pm$ 1.5 \\
\hline
\end{tabular}
\end{table}

\begin{figure*}[h!]
\centering
\subfigure[]{\includegraphics[width=9cm, height=8cm]{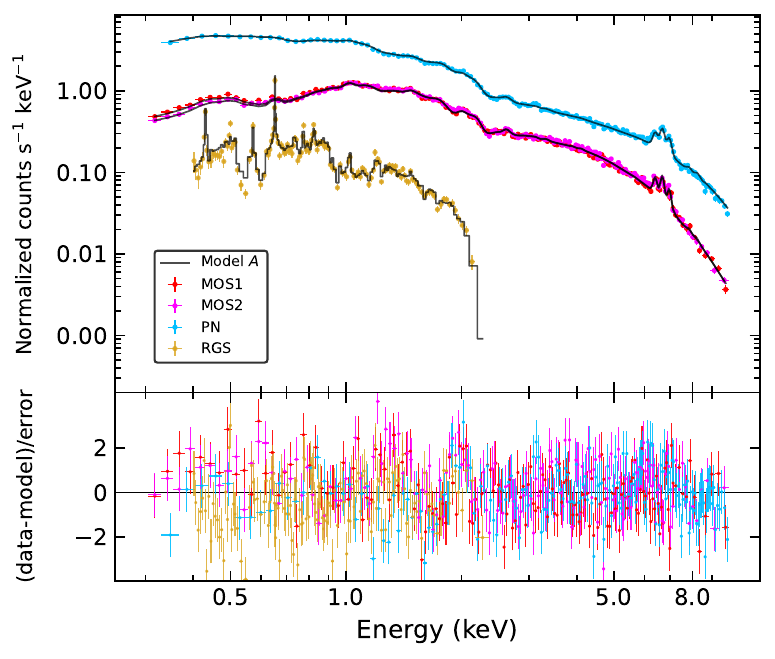} \label{fig:spec_A}}
\subfigure[]{\includegraphics[width=9cm, height=8cm]{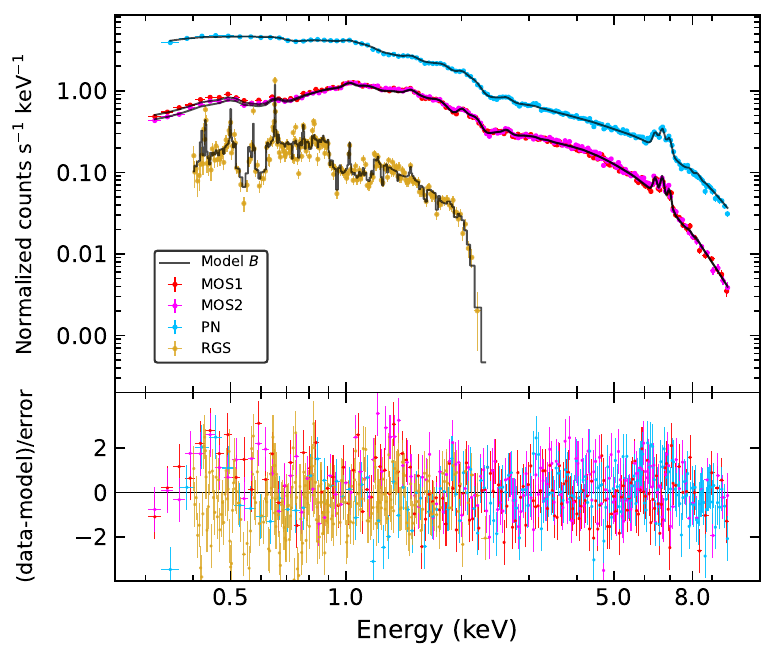}  \label{fig:spec_B}}
\caption{ The MOS1, MOS2, PN, and RGS spectra of J0826 fitted simultaneously with (a) Model $A$ and (b) Model $B$ (see Section \ref{sec_spec} for details). The bottom panel in each figure shows the $\chi^{2}$ contribution of data points for the respective model in terms of residual.}
\label{fig:avg_spec}
\end{figure*}  

\begin{figure}[h!]
\centering
\includegraphics[width=9cm, height=8cm]{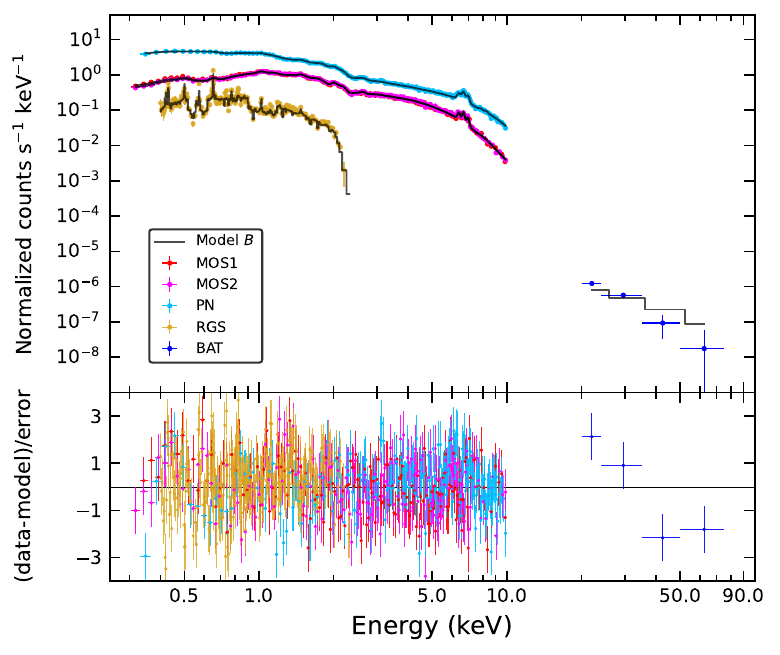}
\caption{ The MOS1, MOS2, PN, RGS, and BAT spectra of J0826 fitted simultaneously with Model $B$ (see Section \ref{sec_spec} for details). The bottom panel shows the $\chi^{2}$ contribution of data points for the respective model in terms of residual.}
\label{fig:avg_spec_global}
\end{figure}

\begin{table*}[ht!]
\renewcommand{\arraystretch}{1.3}
	\centering
\caption {Spectral parameters obtained from fitting the average spectra of J0826 using  model $B$= `\texttt{tbabs$\times$(partcov$\times$tbabs)}$\times$(gsmooth$\times$ $\sum_{i=1}^4 {\rm \texttt{vapec}}_{i}$ + gauss )'.
} \label{tab:spec_modelB}
\begin{tabular}{p{0.5\columnwidth}p{0.5\columnwidth}p{0.35\columnwidth}p{0.35\columnwidth}}
\hline
Model & Parameters  &  without BAT & with BAT\\
\hline
tbabs & N$_{\rm H}$ ($10^{20} \rm ~cm^{-2}$) & $1.4_{-0.4}^{+0.4}$ & $1.7_{-0.4}^{+0.5}$\\

partcov &  cf  & $0.30_{-0.03}^{+0.03}$ & $0.27_{-0.03}^{+0.03}$\\

tbabs & N$_{\rm H,cf}$ ($10^{22} \rm ~cm^{-2}$) & $1.5_{-0.2}^{+0.3} $ &  $1.6_{-0.3}^{+0.4}$\\

gsmooth & $\sigma_{\rm 6 ~keV}$ (eV) & $7.4_{-2.0}^{+2.0}$ & $7.8_{-2.2}^{+2.2}$\\

vapec & T$_{\rm 1}$ (keV) & $0.125_{-0.006}^{+0.007} $ &  $0.122_{-0.006}^{+0.007} $ \\
& Z$_{\rm N}$ (Z$_{\odot}$) & $23.4_{-4.4}^{+5.2} $   & $25.8_{-5.0}^{+6.1} $\\
& Z$_{\rm O}$ (Z$_{\odot}$) &  $7.8_{-1.2}^{+1.4}$ & $8.1_{-1.3}^{+1.5} $\\
& Z$_{\rm Ne}$ (Z$_{\odot}$) & $10.8_{-2.3}^{+2.1}$ & $11.1_{-2.4}^{+2.3} $ \\
& Z$_{\rm Mg}$ (Z$_{\odot}$) & $5.6_{-1.2}^{+1.2} $ & $5.9_{-1.2}^{+1.3} $ \\
& Z$_{\rm Si}$ (Z$_{\odot}$) & $6.9_{-1.2}^{+1.4} $  & $7.2_{-1.3}^{+1.5} $ \\
& Z$_{\rm S}$ (Z$_{\odot}$) & $2.7_{-1.0}^{+1.2} $  & $2.7_{-1.1}^{+1.2} $\\
& Z$_{\rm Fe}$ (Z$_{\odot}$) & $3.2_{-0.3}^{+0.4} $ & $3.2_{-0.3}^{+0.4} $ \\
& N$_{\rm 1}$ (10$^{-4}$)   &  $1.4_{-0.3}^{+0.4} $ &  $1.4_{-0.3}^{+0.4} $\\

vapec & T$_{\rm 2}$ (keV)  &  $0.90_{-0.03}^{+0.04} $  &  $0.90_{-0.03}^{+0.05} $ \\ 
& N$_{\rm 2}$ (10$^{-4}$)   & $1.5_{-0.2}^{+0.3} $ &  $1.4_{-0.2}^{+0.3} $\\ 

vapec & T$_{\rm 3}$ (keV)  &  $4.7_{-0.3}^{+0.3} $  &  $4.7_{-0.3}^{+0.3} $ \\
& N$_{\rm 3}$ (10$^{-3}$)   &  $3.0_{-0.4}^{+0.4} $ &  $2.9_{-0.4}^{+0.4} $ \\

vapec & T$_{\rm 4}$ (keV)  &  $43.2_{-4.9}^{+6.0} $ &  $42.6_{-5.3}^{+5.7} $ \\
& N$_{\rm 4}$ (10$^{-2}$)   &  $1.31_{-0.05}^{+0.05} $ &  $1.27_{-0.06}^{+0.05} $\\

gauss & F$_{\rm 6.4}$ (10$^{-5}$)   &  $3.3_{-0.3}^{+0.3} $  &  $3.3_{-0.3}^{+0.3} $\\
& EW$_{6.4}$ (eV) &  $85.3_{-8.1}^{+8.6} $ &  $86.8_{-3.4}^{+3.4} $ \\

\multirow{1}{*}{Unabsorbed X-ray flux} & F$_{\rm X, unabs}$  (10$^{-11}$ erg cm$^{-2}$ s$^{-1}$)    &    $4.23_{-0.01}^{+0.01}$  & --- \\

\multirow{1}{*}{Absorbed X-ray flux} & F$_{\rm X, abs}$  (10$^{-11}$ erg cm$^{-2}$ s$^{-1}$)    &    $3.856_{-0.009}^{+0.009}$   & --- \\

\multirow{1}{*}{Bolometric flux} & F$_{\rm X, bol}$ (10$^{-10}$ erg cm$^{-2}$ s$^{-1}$)    &    $1.092_{-0.003}^{+0.003}$  &  $1.081_{-0.003}^{+0.003}$\\

\multirow{1}{*}{Bolometric luminosity} & L$_{\rm X, bol}$ (10$^{33}$ erg s$^{-1}$)    &    $1.759_{-0.004}^{+0.004}$  &   $1.739_{-0.004}^{+0.004}$   \\
& $\chi_{\nu}^2$ (dof)   &  1.36 (1237)       & 1.23 (1237)  \\
\hline
\end{tabular}

\bigskip
\textbf{Note.} $\sigma_{\rm 6 ~keV}$ is the line width of gaussian lines at 6 keV. The abundances of the other three vapec components were tied to the first component.
\end{table*}

\section{Spectral analysis} \label{sec_spec}
The background-subtracted X-ray spectra of J0826  were analyzed in the energy range 0.3-10.0 keV using \textsc{xspec} version 12.12.0 \citep{1996ASPC..101...17A, 2001ASPC..238..415D}. Typically, the X-ray spectrum of IPs is characterized by the presence of multi-temperature thermal plasma emission components along with neutral total and partial covering absorbers. The continuum along with the H- and He-like ion emission lines of N, O, and Ne and strong iron emission line features at 6.4 keV (Fe K$\alpha$), 6.7 keV (Fe XXV), and 6.95 keV (Fe XXVI) are clearly seen in the spectra of J0826 (see Fig. \ref{fig:avg_spec}). Therefore, to mimic the X-ray spectra of J0826, we used various model components in combination with each other available in \textsc{xspec}. The model \texttt{tbabs} is used to account for the X-ray absorption by the interstellar medium along the line of sight, sometimes in combination with convolution model \texttt{partcov}, which represents a partial covering absorber.  The only parameter of \texttt{partcov} provides the value of the covering fraction (cf) of the absorber. To represent emission from
collisionally ionized diffuse gas due to accretion, we used the model \texttt{apec}. It has three parameters: plasma temperature (T in keV), metal abundance relative to the solar value (Z), and normalization (N in units of cm$^{-5}$). The normalization of the \texttt{apec} component is the volume emission measure and related by  $\frac{10^{-14}}{4\pi \rm D^2}$ $\int \rm n_e n_H dV$, where D is the distance to the source, $\rm n_{e}$ and $\rm n_{H}$ are the electron density and hydrogen density, respectively, and V is the emitting volume.

\par We first used a simple model consisting of \texttt{tbabs} and adding from one to three \texttt{apec} components to represent the continuum plasma emission spectrum. The fit with three \texttt{apec} components was not able to account for emission line features of O, Ne, and N observed in the combined first order RGS spectrum. Therefore we added five \texttt{Gaussian} components at 0.43 keV, 0.50 keV, 0.57 keV, 0.65 keV, and 1.02 keV, which were identified as N \rn{6}, N \rn{7}, O \rn{7},  O \rn{8}, and Ne \rn{10}, respectively. The \texttt{Gaussian} component has three parameters: line energy (in keV), line width ($\sigma$), and line ﬂux (F) in terms of photons $\rm cm^{-2} \rm s^{-1}$, respectively. A further \texttt{Gaussian} component was used for the fluorescent iron line at 6.4 keV. However, this fit was poor with a  $\chi_\nu^2$ value of $\sim$1.7. The \texttt{partcov}  was then used along with \texttt{tbabs} due to the presence of residuals at low energies. 
\noindent Thus, the model \texttt{tbabs$\times$(partcov$\times$tbabs)}$\times$($\sum_{i=1}^3 {\rm \texttt{apec}}_{i}$ + $\sum_{i=1}^6 {\rm \texttt{gauss}}_{i}$ ) slightly improved the fit with a $\chi_\nu^2$ value of 1.6. Further, adding a fourth \texttt{apec} component in this model remarkably improved the fit, with $\chi_\nu^2$ value of 1.30. We employed the `F-test' to verify the necessity of the fourth \texttt{apec} component and found an F-statistic value of 40.6 with a null hypothesis probability of 5.4 $\times$ 10$^{-25}$. In this way, we define our model $A$ as  `\texttt{tbabs$\times$(partcov$\times$tbabs)}$\times$($\sum_{i=1}^4 {\rm \texttt{apec}}_{i}$ + $\sum_{i=1}^6 {\rm \texttt{gauss}}_{i}$ )'.  With model $A$, the metal abundance was found to be higher than solar with a value of $2.7\pm0.2$ (see Table \ref{tab:spec_modelA} for spectral parameters obtained from this fitting). Additionally, we investigated whether model $A$ excluding \texttt{Gaussian} components could constrain the emission line features of various lines observed in the RGS spectrum. However, our analysis revealed that it was unable to do so.

\par Since we found a higher value than solar for the metal abundance with model $A$, we replaced \texttt{apec} with \texttt{vapec} to better understand the effect of individual abundances. The \texttt{vapec} model is the same as \texttt{apec}, but it allows modification of element abundances, such as He, C, N, O, Ne, Mg, Al, Si, S, Ar, Ca, Fe, and Ni. However, replacing \texttt{apec} with \texttt{vapec} and omitting \texttt{Gaussian} components in model $A$ also resulted in the inadequate fitting of the emission line features. Therefore, we used a convolution model \texttt{gsmooth} with \texttt{vapec}, which broadens the emission lines. We fixed the power law index of \texttt{gsmooth} to 1 so that all the lines are broadened by a constant velocity. In this manner, model $B$ = `\texttt{tbabs$\times$(partcov$\times$tbabs)}$\times$(gsmooth$\times$ $\sum_{i=1}^4 {\rm \texttt{vapec}}_{i}$ + gauss )' was found to be the best-fit model with a $\chi_\nu^2$ value of 1.36 because it was able to represent various emission line features present in the spectra, which the \texttt{apec} components could not represent. In model $B$, the N, O, Ne, Mg, Si, S, and Fe abundances were treated as free parameters. Meanwhile, the abundances of He, C, Al, Ar, Ca, and Ni were fixed to the solar values, as these elements do not contribute within the covered energy range and therefore, their values could not be constrained when left free.  In both models, we have adopted abundances derived by \cite{2000ApJ...542..914W}.  In order to further study the X-ray spectrum over a wider range, we have also used the average spectrum from the BAT 105-month catalogue \citep{2018ApJS..235....4O}\footnote{\url{https://swift.gsfc.nasa.gov/results/bs105mon/417}} along with the relevant response files\footnote{\url{https://swift.gsfc.nasa.gov/results/bs105mon/}}. We have performed the combined spectral fitting in the 0.3-75.0 keV energy range using the best-fit model $B$. The unabsorbed bolometric flux in the 0.001–100.0 keV energy band was also calculated by incorporating the \texttt{cflux} component in models $A$ and $B$. The spectral parameters derived from the simultaneous fitting to PN, MOS, RGS, and BAT spectra using model $B$ together with the 90\% confidence limit for a single parameter are given in Table \ref{tab:spec_modelB}.

\begin{table}[h!]
\renewcommand{\arraystretch}{1.3}
	\centering
\caption {Spectral parameters as derived from fitting the EPIC MOS and PN spectra at maximum and minimum phases of \pone ~using Model $B$. } \label{tab:phase_resolved_p1_modelB}
\begin{tabular}{p{0.28\columnwidth}p{0.28\columnwidth}p{0.28\columnwidth}}
\hline
Parameters  &  Pulse Maximum & Pulse Minimum \\
\hline

cf  & $0.30_{-0.02}^{+0.02}$ & $0.28_{-0.03}^{+0.03}$\\
N$_{\rm H,cf}$ ($10^{22} \rm ~cm^{-2}$) & $1.4_{-0.1}^{+0.1} $  &  $1.6_{-0.2}^{+0.2} $ \\

N$_{\rm 1}$ (10$^{-4}$)   &  $1.8_{-0.2}^{+0.2} $ & $1.2_{-0.2}^{+0.2} $ \\

N$_{\rm 2}$ (10$^{-4}$)   & $1.6_{-0.2}^{+0.2} $ & $1.5_{-0.2}^{+0.2} $ \\ 

N$_{\rm 3}$ (10$^{-3}$)   &  $2.9_{-0.2}^{+0.2} $ & $3.0_{-0.3}^{+0.3} $\\

N$_{\rm 4}$ (10$^{-2}$)   &  $1.37_{-0.02}^{+0.02} $ &   $1.20_{-0.03}^{+0.03} $\\

F$_{\rm 6.4}$ (10$^{-5}$)   &  $3.5_{-0.4}^{+0.4} $ &  $3.0_{-0.5}^{+0.5} $\\

F$_{\rm X}$ (10$^{-11}$erg~cm$^{-2}$~s$^{-1}$) & $4.40_{-0.01}^{+0.01}$ & $3.96_{-0.02}^{+0.02} $\\ 

$\chi_\nu^2$ (dof)   &  1.11 (523)      &   1.12 (500)  \\
\hline
\end{tabular}
\bigskip
\begin{tablenotes}
	\small {
	\item{\textbf{Note.}} F$_{\rm X}$ is the unabsorbed X-ray flux in the 0.3-10.0 keV energy range.}
    \end{tablenotes}
\end{table}

\begin{table}[h!]
\renewcommand{\arraystretch}{1.3}
\centering
\caption {Spectral parameters as derived from fitting the EPIC MOS and PN spectra at maximum and minimum phases of \pthree ~using Model $B$. } \label{tab:phase_resolved_p3_modelB}
\begin{tabular}{p{0.28\columnwidth}p{0.28\columnwidth}p{0.28\columnwidth}}
\hline
Parameters  &  Pulse Maximum & Pulse Minimum \\
\hline

cf  & $0.30_{-0.02}^{+0.02}$ & $0.28_{-0.02}^{+0.02}$\\
N$_{\rm H,cf}$ ($10^{22} \rm ~cm^{-2}$) & $1.6_{-0.2}^{+0.2} $  &  $1.4_{-0.2}^{+0.2} $ \\

N$_{\rm 1}$ (10$^{-4}$)   &  $1.6_{-0.2}^{+0.2} $ & $1.4_{-0.2}^{+0.2} $ \\

N$_{\rm 2}$ (10$^{-4}$)   & $1.5_{-0.2}^{+0.2} $  & $1.5_{-0.2}^{+0.2} $ \\

N$_{\rm 3}$ (10$^{-3}$)   &  $3.2_{-0.2}^{+0.2} $ & $2.8_{-0.2}^{+0.2} $\\

N$_{\rm 4}$ (10$^{-2}$)   &  $1.34_{-0.02}^{+0.02} $ &   $1.26_{-0.02}^{+0.02} $\\

F$_{\rm 6.4}$ (10$^{-5}$)   &  $4.0_{-0.5}^{+0.5} $ &  $2.7_{-0.4}^{+0.4} $\\

F$_{\rm X}$ (10$^{-11}$erg~cm$^{-2}$~s$^{-1}$) & $4.40_{-0.02}^{+0.02}$ & $4.05_{-0.01}^{+0.01} $\\ 

$\chi_\nu^2$ (dof)   &  1.03 (514)      &   1.32 (513)  \\
\hline
\end{tabular}

\bigskip
\begin{tablenotes}
\small {\item{\textbf{Note.}} F$_{\rm X}$ is the unabsorbed X-ray flux in the 0.3-10.0 keV energy range.}
\end{tablenotes}

\end{table}

\subsection{Phase-resolved spectroscopy}
To explore the possible changes in the spectral parameters along the minimum and maximum phase ranges of \pone ~and \pthree, we extracted the MOS and PN spectra at pulse maximum and minimum identified on the total 0.2-12.0 keV light curve. The maximum and minimum phase ranges were identified as 0.7-1.3 and 0.3-0.7 for \pone ~and 0.6-1.1 and 0.1-0.6 for \pthree, respectively. The spectral fitting was performed using the best-fit model $B$ and fixing all other parameters to the value found for the average spectrum except for the covering fraction, N$_{\rm H,cf}$, and normalizations of thermal and Gaussian components. The spectral parameters using the model $B$ together with the 90\% confidence limit for a single parameter are given in Tables \ref{tab:phase_resolved_p1_modelB} and \ref{tab:phase_resolved_p3_modelB}.  No substantial change in the parameters is found for both periods; however, there are changes in the normalization of the hotter \texttt{vapec} component N$_{4}$.  The value of N$_{4}$ is larger at pulse maximum than at pulse minimum. On the other hand, these values are within 3-5$\sigma$ consistent, which indicates that N$_{4}$ shows a hint of variability at \pone ~and P$_{3}$ periodicities. This might imply that the hotter component is the one that drives the variability at the two periods due to changes in the projected area.

\subsection{Estimation of the WD mass}
In order to derive an estimate of the WD mass, we used the post-shock region (PSR) model of \citep{2016A&A...591A..35S, 2019MNRAS.482.3622S}, also known as \texttt{ipolar} in \textsc{xspec}. It has three parameters: fall height, WD mass, and normalization. The fall height is the height from which the accretion flow starts to fall freely, i.e. magnetospheric radius (R$_{\rm m}$). The aperiodic power spectrum shows no presence of break frequency; therefore by making an assumption that R$_{\rm m}$ is equal to the co-rotation radius \citep{2019MNRAS.482.3622S, 2020MNRAS.498.3457S}, we have set the fall height parameter to be linked with the WD mass, as also done in \cite{2022MNRAS.511.4582C} and \cite{2023MNRAS.523.4520T}. To avoid emission line effects, we fitted the spectra in the energy range 8.0-75.0 keV using the PSR model as  `\texttt{tbabs$\times$(partcov$\times$tbabs)$\times$(atable$\{$ipolar.fits$\}$)'}, which resulted in a $\chi_\nu^2$ value of 0.75. The covering fraction and densities were set to the values found in the average spectral fitting.  Hence, we derived the WD mass and normalization to be 0.7$_{-0.2}^{+0.2}$ M$_{\odot}$ and 7.0$_{-4.5}^{+10.1}$ $\times$10$^{-27}$, respectively. Moreover, the fall height is estimated to be 47.3 R$_{\rm WD}$ (=36.5 $\times$ 10$^{4}$ km), in which we have considered our candidate spin period value of $\sim$4588 s and above mentioned value of WD mass along with the WD mass-radius relationship of \cite{1972ApJ...175..417N}. We also derived the WD mass of 0.87$_{-0.05}^{+0.06}$  M$_{\odot}$ by adopting the maximum temperature of $42.6_{-5.3}^{+5.7}$ keV found in the average spectral fitting as the shock temperature. The shock temperature (T$_{\rm s}$) is related to WD mass as kT$_{\rm s}$ = $\frac{\rm 3GM_{\rm WD} \mu m_{\rm H}}{\rm 8 R_{\rm WD}}$, where k is the Boltzmann constant, G is the gravitational constant, $\mu$ is the mean molecular weight, and m$_{\rm H}$ is the mass of the hydrogen atom. The discrepancy with \citet{2012A&A...545A.101P} is clearly due to the underestimated value of shock temperature derived using the lower quality data from Swift/XRT. Moreover, our derived value of the WD mass using two different methods is well within a 1$\sigma$ consistent with each other and also in good agreement with the mean value found for CVs by \cite{2000MNRAS.314..403R} and \citep{2022MNRAS.510.6110P}.

\section{Discussion}
Our XMM-Newton observation of J0826 has revealed the indications of magnetic accretion in this system, which are described in detail in upcoming sections.

\subsection{The variability properties} \label{sec5.1} 
The timing analyses revealed the presence of three X-ray peaks in the periodogram at \pone=14103 $\pm$ 149 s, P$_{2}$=5604 $\pm$ 34 s, and \pthree=4588 $\pm$ 25 s. These variabilities are found to be significant at the 95$\%$ global significance level.  The same variability is also found in the optical data.  The longer period at $\sim$3.92\,h would naturally be ascribed to the orbital period of the system. However, we have a suspicion that P$_{1}$ might be the half of the orbital period (\po), P$_{3}$ the spin period (\ps), and P$_{2}$ the beat between spin and orbital periods (\pb). The beat frequency ($\omega-\Omega$) arises in IPs due to the flipping of the accretion stream twice between the magnetic poles as  WD rotates with respect to the binary frame. If we consider P$_{1}$ as the orbital period and P$_{3}$ as the candidate spin period, then the estimated value of \pb ~comes out to be 6800 $\pm$ 65 s, which is beyond the 3$\sigma$ significance level with the observed value of P$_{2}$. Whereas, considering P$_{1}$ as half of ~\po, the beat period is estimated to be 5479 $\pm$ 37 s. This value is well within a 2$\sigma$ significance level with the observed value of P$_{2}$, which represents our candidate beat period. Therefore, to reconcile the presence of the other two periodicities, if \pthree ~represents the spin and $\rm P_2$ the orbital negative sideband ($\omega - \Omega$), the period P$_{1}$ is half of the actual orbital period which would result to be $\sim$7.84\,h.   Such a long period would locate J0826 near the far end of the IP orbital period distribution \citep{2020AdSpR..66.1209D} and close to V2069\,Cyg. The significance of these three peaks in the power spectra is high enough to consider them as true, suggesting an IP classification.  If our candidate spin and beat periods are true periods, then J0826 accretes via a combination of disc and stream (disc-overflow accretion), with an equal fraction occurring via disc and stream, as evident from the equal values of fractional amplitudes obtained for \ps and \pb.  J0826 would also have a degree of asynchronism  P$_{\omega}$/P$_{\Omega}$ of $\sim$0.16, comparatively larger than other IPs in the orbital period range of $\sim$7-8 h, e.g. V902 Mon (0.08), IGR J08390-4833 (0.05), Swift J0939.7-3224 (0.09), CXOGBS J174954.5-294335 (0.02), V2069 Cyg (0.03), and 1RXS J213344.1+510725 (0.02). Hence J0826 would be located in the spin-orbit period plane among the slowest WD rotators for its orbital period.

\par
\begin{figure}[htb!]
\centering
\includegraphics[width=9cm, height=8cm]{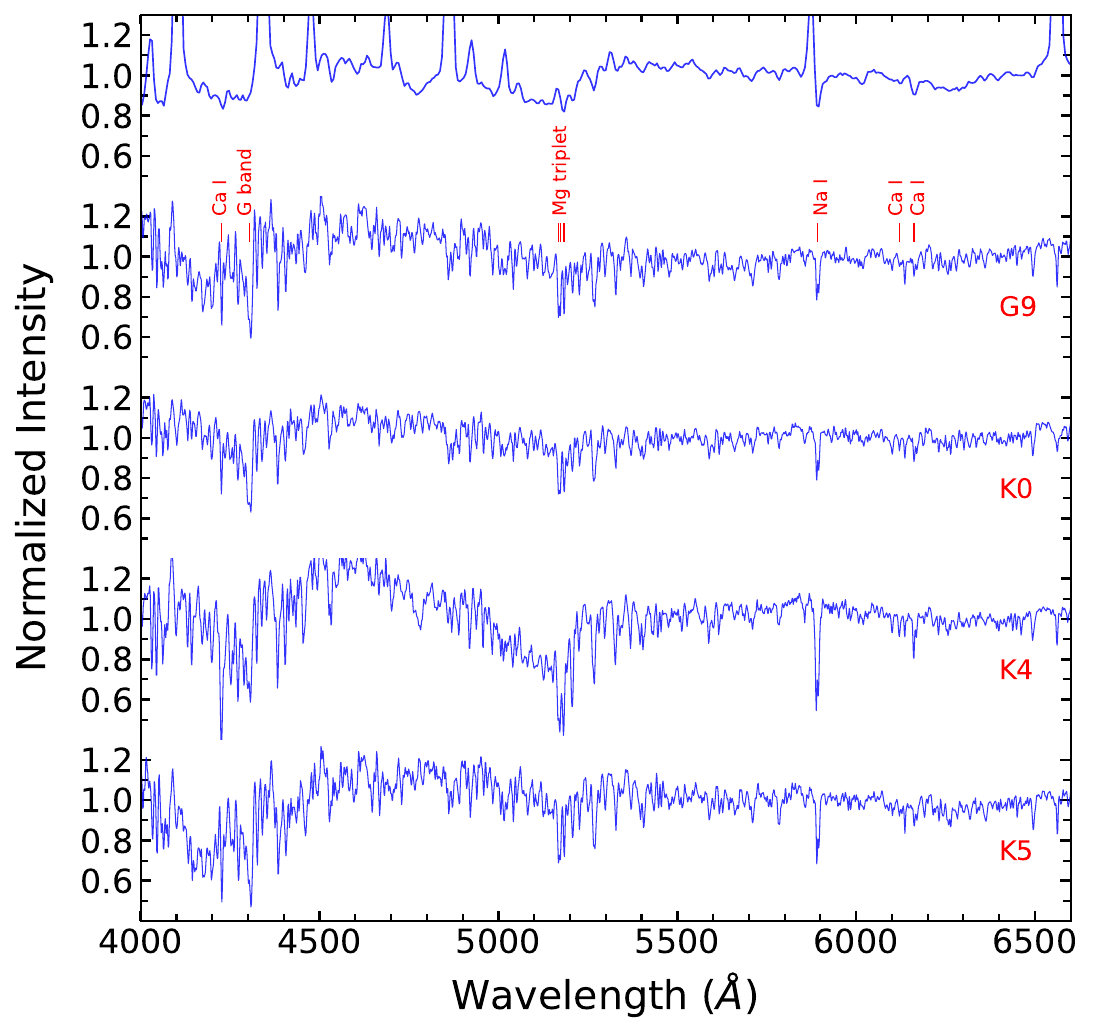}
\caption{Comparison of optical spectrum of \cite{2012A&A...545A.101P} with stellar spectra of spectral type G9, K0, K4, and K5. }
\label{fig:parisi}
\end{figure}

 Considering the candidate orbital period value of 7.84 h, the mean density of the secondary is estimated to be $\sim$1.7 g cm$^{-3}$ \citep[see][for formulae]{1995cvs..book.....W}, which falls in the range of mean densities of a main-sequence star of G-K type \citep{1976asqu.book.....A}. Furthermore, we retrieved the optical spectrum by \cite{2012A&A...545A.101P}\footnote{The spectrum is publicly available at CDS at  \url{http://cdsarc.u-strasbg.fr/viz-bin/qcat?J/A+A/545/A101}.} and found that the flux calibrated spectrum is quite off with respect to the optical photometry (about 5 times fainter),  indicating either a non-optimal flux calibration or that the short (900 s) exposure caught the source at a lower flux given its variability. We, therefore, normalized the spectrum to the continuum and identified a few weak absorption features ascribed to Ca\,I 4226 \AA, weak 6122 \AA, 6162 \AA, Na\,I 5893 \AA, Mg triplet (5167 \AA, 5173 \AA, and 5183 \AA) and possibly the G-band at 4310 \AA. The relatively low resolution optical spectrum does not allow to obtain a precise identification of the companion spectral type. Given the low-resolution spectrum, we only performed comparisons with stellar spectra from \cite{1984ApJS...56..257J}, encompassing spectral types G9, K0, K4, and K5 (see, Fig. \ref{fig:parisi}). A K6 star is not favoured due to the lack of strong molecular bands. The Ca\,I 6122 \AA ~is very weak, indicating a spectral type not earlier than K3. Therefore, based on the limited information, we adopt a K4-K5 spectral type for secondary.  On the other hand, if the true orbital period is P$_{1}$,  the secondary star would be an M3.6 star \citep{2011ApJS..194...28K}, hence of a much later spectral type to be consistent with the absorption features observed in the optical spectrum. These findings suggest that J0826 belongs to the category of long orbital period systems with evolved donors, which is consistent with the suggestions given by \cite{1998A&A...339..518B}, \cite{2000MNRAS.318..354B}, and \cite{2003MNRAS.340.1214P}. However, the identification of the true orbital period would reside in the study of radial velocities in the optical. 

\par Unlike the majority of IPs, the phase-folded analyses reveal that modulations at our proposed spin period and suspected half of the orbital period are energy independent, which suggests that these modulations are only due to the changes in the projected emitting area and not due to the photoelectric absorption in the accretion curtain. The 2$\Omega$ frequency, along with $\Omega$ frequency, has also been seen in $\sim$62\% of the sample studied by \cite{2005A&A...439..213P} in the X-ray wavelength.  Furthermore, optical modulations at our candidate half of the orbital period and spin period are in phase with the X-ray modulations, consistent with patterns seen in most IPs. This indicates that the optical spin pulsation originates in the magnetically confined accretion flow.

\subsection{The post-shock region} \label{sec5.2}
The average X-ray spectral analyses reveal that the X-ray post-shock emitting region exhibits a multi-temperature structure with temperatures  $0.122_{-0.006}^{+0.007}$ keV, $0.90_{-0.03}^{+0.05}$ keV,  $4.7_{-0.3}^{+0.3}$ keV, and $42.6_{-5.3}^{+5.7}$ keV. The spectrum indicates absorption by a total absorber with N$_{\rm H}$ of $\sim$1.7 $\times$ 10$^{20}$ cm$^{-2}$ and a local absorber with N$_{\rm H}$ of $\sim$1.6 $\times$ 10$^{22}$ cm$^{-2}$. The latter covers $\sim$27$\%$ of the X-ray source. The column density of the total absorber is much lower than the total interstellar value in the direction of the source, i.e. $1.2\times10^{21}$ cm$^{-2}$ and consistent with the close distance of 366\,pc derived from the Gaia parallax. The partial covering absorber density is lower than typical for IPs but still within the plausible range \citep{2017PASP..129f2001M}. The moderately weak equivalent width (EW) of the fluorescent iron line at 6.4 keV indicates that reflection from the WD surface does not play important role in J0826, which also justifies the redundancy of this component in the spectral fits. 

\par
\begin{figure}[htb!]
\centering
\includegraphics[width=9cm, height=6cm]{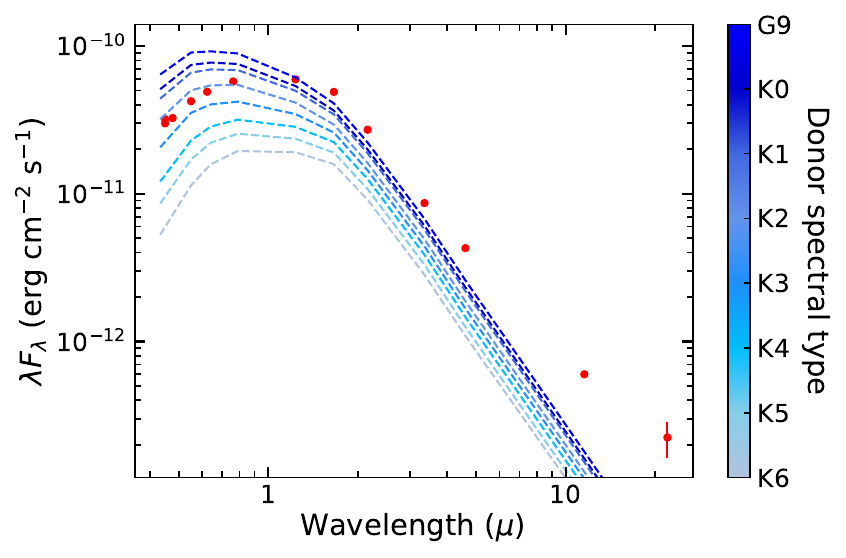}
\caption{The spectral energy distribution of J0826. The red filled circles represent the photometric data, while the dashed curves represent the donors of spectral type ranging from G9 to K6. }
\label{fig:sed}
\end{figure}

 A lower limit to the accretion luminosity can be derived using the unabsorbed bolometric luminosity derived from the spectral fitting $\rm L_{X,bol} = $ 1.7 $\times$ 10$^{33}$ erg s$^{-1}$ adopting the Gaia distance of 366\,pc. However, to obtain a more reliable value of the accretion luminosity, we also consider the optical-nIR emission using the OM B-band and APASS B, V, g, r band measures, the 2MASS J, H, K and WISE W1, W2, W3, and W4 photometry\footnote{The photometric measures are retrieved using  VizieR access catalogue tool at CDS.}, corrected for interstellar extinction E(B-V)=0.02 estimated from the hydrogen column density found from X-ray spectral fit using the relation of \citep{2009MNRAS.400.2050G}.  Here we note that J0826 when observed by XMM-Newton OM  has been found at about the same level as in the APASS and consistently in the All-Sky Automated Survey for Supernovae \citep[ASAS-SN;][]{2014ApJ...788...48S}, which allows us to bridge the X-ray and ground-based photometry. Given the observed dereddened magnitudes of J0826 from B (15.1\,mag) to nIR in the K-band (11.3\,mag), only a donor with spectral type K2 or later located at 366\,pc can account for the observed magnitudes. This indicates that the contribution of accretion at optical-nIR wavelengths is not dominant. Assuming a K4-K5 star, which is also evident from our earlier analysis (see Section \ref{sec5.1}), the excess of flux occurs in the optical-nIR bands (see, Fig. \ref{fig:sed}), which results to be F$\rm _{excess}$ $\sim$4.8$-$5.6 $\times$ 10$^{-11}$ erg\,cm$^{-2}$\,s$^{-1}$, corresponding to a luminosity of L $\sim$7.7$-$9.0 $\times$ 10$^{32}$ erg\,s$^{-1}$. Therefore, we estimate an accretion luminosity as $\rm L_{acc} = L_{X,bol} + L = 2.47-2.6\times 10^{33}\, erg\,s^{-1}$. Using L$_{\rm acc}$ = GM$_{\rm WD}$ $\dot{\rm M}$/R$_{\rm WD}$ and adopting the WD mass value of 0.7 M$_{\odot}$, we derive a mass accretion rate of 2.2$-$2.3 $\times$ 10$^{-10}$ M$_{\odot}$ yr$^{-1}$.  This value is at least two orders of magnitude lower than expected from magnetic braking \citep[4.1 $\times$ 10$^{-8}$ M$_{\odot}$ yr$^{-1}$;][]{1989ApJ...342.1019M} for an orbital period of $\sim$7.84 h.  However, \cite{2000MNRAS.318..354B} have shown that a low mass transfer rate ($\leq$5 $\times$ 10$^{-9}$ M$_{\odot}$ yr$^{-1}$) can be expected for orbital periods $\geq$ 6 h for evolved donors. 

\par The striking features of J0826 are substantial suprasolar abundance of N ($25.8_{-5.0}^{+6.1}$), O ($8.1_{-1.3}^{+1.5}$), Ne ($11.1_{-2.4}^{+2.3}$), Mg ($5.9_{-1.2}^{+1.3}$), Si ($7.2_{-1.3}^{+1.5}$), S ($2.7_{-1.1}^{+1.2}$), and Fe ($3.2_{-0.3}^{+0.4}$). On the basis of anomalous UV line flux ratios obtained through UV spectroscopy, \cite{2003ApJ...594..443G} estimated that at least 10-15$\%$ fraction of CVs might display large abundance anomalies. This might be attributed to the fact that these CVs might have gone through the thermal timescale mass transfer (TTMT) phase, in which the CV originally contained a donor more massive than the WD \citep{2002MNRAS.337.1105S, 2003MNRAS.340.1214P, 2023MNRAS.523..305T}. The suprasolar abundances with an indication of a long orbital period and evolved donor, collectively suggest that J0826 likely underwent TTMT \citep{2004ASPC..315....8S}. Similar indications were also proposed for V1082 Sgr by \cite{2013MNRAS.435.2822B}. Therefore, accurate photometry and polarimetry in optical wavelengths and UV spectroscopy will play an essential role in understanding the true nature and evolutionary status of this CV.

\section{Conclusions}
We have presented the first long X-ray observation of J0826 with XMM-Newton, which gives an indication of this CV belonging to the IP class. Here we summarize the main results:

\begin{itemize}
    \item The variability in X-rays and optical wavelengths shows modulations at short and long timescales. The long timescale variability ($\sim$14000 s) could be attributed to half of the true orbital period. Whereas $\sim$4600 s and $\sim$5600 s variabilities might correspond to spin and beat periods of the system, respectively.  The simultaneous presence of spin and beat periods may suggest that J0826 accretes via disc-overflow.

    \item The putative spin and orbital variabilities are energy independent, suggesting that these are due to the changes in the projected emitting area.

    \item The post-shock region is comprised of a multi-temperature structure with a maximum temperature of about 43 keV, from which
    we derive a WD mass in the range of 0.82 - 0.93 M$_{\odot}$,
    which is in good agreement with the WD mass of 0.7 $\pm$ 0.2 M$_{\odot}$ derived using the PSR model.
    
    \item The X-ray spectrum remarkably shows the presence of suprasolar abundances, which together with the indication of a long orbital period and an evolved donor, collectively suggest that J0826 probably underwent a TTMT phase.

\end{itemize}

\begin{acknowledgements}
We thank the anonymous referee for providing valuable comments and
suggestions that led to a significant improvement in the quality of
the paper. This work is based on observations obtained with XMM-Newton, an ESA science mission with instruments and contributions directly funded by ESA Member States and NASA. We acknowledge the use of public data from the Swift data archive. This work has made use of the VizieR catalogue access tool, CDS, Strasbourg, France (DOI : 10.26093/cds/vizier) as well as of the European Space Agency (ESA) space mission Gaia data which are processed by the Gaia Data Processing and Analysis Consortium (DPAC). Funding for the DPAC is provided by national institutions, in particular, the institutions participating in the Gaia MultiLateral Agreement (MLA). The Gaia mission website is \url{https://www.cosmos.esa.int/gaia}. NR extends gratitude to Dr Arti Joshi and Mr Rahul Batra for helpful discussions. NM and DdM acknowledge financial support through ASI-INAF agreement 2017-14-H.0 (PI: T. Belloni).
\end{acknowledgements}

\bibliographystyle{aa}
\bibliography{ref.bib}

\begin{thebibliography}{47}
\expandafter\ifx\csname natexlab\endcsname\relax\def\natexlab#1{#1}\fi

\bibitem[{{Aizu}(1973)}]{1973PThPh..49.1184A}
{Aizu}, K. 1973, Progress of Theoretical Physics, 49, 1184

\bibitem[{{Allen}(1976)}]{1976asqu.book.....A}
{Allen}, C.~W. 1976, {Astrophysical Quantities}

\bibitem[{{Arnaud}(1996)}]{1996ASPC..101...17A}
{Arnaud}, K.~A. 1996, in Astronomical Society of the Pacific Conference Series,
  Vol. 101, Astronomical Data Analysis Software and Systems V, ed. G.~H.
  {Jacoby} \& J.~{Barnes}, 17

\bibitem[{{Bailer-Jones} {et~al.}(2021){Bailer-Jones}, {Rybizki}, {Fouesneau},
  {Demleitner}, \& {Andrae}}]{2021AJ....161..147B}
{Bailer-Jones}, C.~A.~L., {Rybizki}, J., {Fouesneau}, M., {Demleitner}, M., \&
  {Andrae}, R. 2021, AJ, 161, 147

\bibitem[{{Baraffe} \& {Kolb}(2000)}]{2000MNRAS.318..354B}
{Baraffe}, I. \& {Kolb}, U. 2000, \mnras, 318, 354

\bibitem[{{Bernardini} {et~al.}(2013){Bernardini}, {de Martino}, {Mukai},
  {Falanga}, {Andruchow}, {Bonnet-Bidaud}, {Masetti}, {Buitrago}, {Mouchet}, \&
  {Tovmassian}}]{2013MNRAS.435.2822B}
{Bernardini}, F., {de Martino}, D., {Mukai}, K., {et~al.} 2013, \mnras, 435,
  2822

\bibitem[{{Beuermann} {et~al.}(1998){Beuermann}, {Baraffe}, {Kolb}, \&
  {Weichhold}}]{1998A&A...339..518B}
{Beuermann}, K., {Baraffe}, I., {Kolb}, U., \& {Weichhold}, M. 1998, \aap, 339,
  518

\bibitem[{{Coughenour} {et~al.}(2022){Coughenour}, {Tomsick}, {Shaw}, {Mukai},
  {Clavel}, {Hare}, {Krivonos}, \& {Fornasini}}]{2022MNRAS.511.4582C}
{Coughenour}, B.~M., {Tomsick}, J.~A., {Shaw}, A.~W., {et~al.} 2022, \mnras,
  511, 4582

\bibitem[{{Cusumano} {et~al.}(2010){Cusumano}, {La Parola}, {Segreto},
  {Mangano}, {Ferrigno}, {Maselli}, {Romano}, {Mineo}, {Sbarufatti}, {Campana},
  {Chincarini}, {Giommi}, {Masetti}, {Moretti}, \&
  {Tagliaferri}}]{2010A&A...510A..48C}
{Cusumano}, G., {La Parola}, V., {Segreto}, A., {et~al.} 2010, \aap, 510, A48

\bibitem[{{de Martino} {et~al.}(2020){de Martino}, {Bernardini}, {Mukai},
  {Falanga}, \& {Masetti}}]{2020AdSpR..66.1209D}
{de Martino}, D., {Bernardini}, F., {Mukai}, K., {Falanga}, M., \& {Masetti},
  N. 2020, Advances in Space Research, 66, 1209

\bibitem[{{den Herder} {et~al.}(2001){den Herder}, {Brinkman}, {Kahn},
  {Branduardi-Raymont}, {Thomsen}, {Aarts}, {Audard}, {Bixler}, {den Boggende},
  {Cottam}, {Decker}, {Dubbeldam}, {Erd}, {Goulooze}, {G{\"u}del}, {Guttridge},
  {Hailey}, {Janabi}, {Kaastra}, {de Korte}, {van Leeuwen}, {Mauche},
  {McCalden}, {Mewe}, {Naber}, {Paerels}, {Peterson}, {Rasmussen}, {Rees},
  {Sakelliou}, {Sako}, {Spodek}, {Stern}, {Tamura}, {Tandy}, {de Vries},
  {Welch}, \& {Zehnder}}]{2001A&A...365L...7D}
{den Herder}, J.~W., {Brinkman}, A.~C., {Kahn}, S.~M., {et~al.} 2001, A\&A,
  365, L7

\bibitem[{{Dorman} \& {Arnaud}(2001)}]{2001ASPC..238..415D}
{Dorman}, B. \& {Arnaud}, K.~A. 2001, in Astronomical Society of the Pacific
  Conference Series, Vol. 238, Astronomical Data Analysis Software and Systems
  X, ed. J.~{Harnden}, F.~R., F.~A. {Primini}, \& H.~E. {Payne}, 415

\bibitem[{{G{\"a}nsicke} {et~al.}(2003){G{\"a}nsicke}, {Szkody}, {de Martino},
  {Beuermann}, {Long}, {Sion}, {Knigge}, {Marsh}, \&
  {Hubeny}}]{2003ApJ...594..443G}
{G{\"a}nsicke}, B.~T., {Szkody}, P., {de Martino}, D., {et~al.} 2003, \apj,
  594, 443

\bibitem[{{G{\"u}ver} \& {{\"O}zel}(2009)}]{2009MNRAS.400.2050G}
{G{\"u}ver}, T. \& {{\"O}zel}, F. 2009, \mnras, 400, 2050

\bibitem[{{Hameury} {et~al.}(1986){Hameury}, {King}, \&
  {Lasota}}]{1986MNRAS.218..695H}
{Hameury}, J.~M., {King}, A.~R., \& {Lasota}, J.~P. 1986, MNRAS, 218, 695

\bibitem[{{Hellier}(1995)}]{1995ASPC...85..185H}
{Hellier}, C. 1995, in Astronomical Society of the Pacific Conference Series,
  Vol.~85, Magnetic Cataclysmic Variables, ed. D.~A.~H. {Buckley} \&
  B.~{Warner}, 185

\bibitem[{{Jacoby} {et~al.}(1984){Jacoby}, {Hunter}, \&
  {Christian}}]{1984ApJS...56..257J}
{Jacoby}, G.~H., {Hunter}, D.~A., \& {Christian}, C.~A. 1984, ApjS, 56, 257

\bibitem[{{Jansen} {et~al.}(2001){Jansen}, {Lumb}, {Altieri}, {Clavel}, {Ehle},
  {Erd}, {Gabriel}, {Guainazzi}, {Gondoin}, {Much}, {Munoz}, {Santos},
  {Schartel}, {Texier}, \& {Vacanti}}]{2001A&A...365L...1J}
{Jansen}, F., {Lumb}, D., {Altieri}, B., {et~al.} 2001, A\&A, 365, L1

\bibitem[{{Knigge} {et~al.}(2011){Knigge}, {Baraffe}, \&
  {Patterson}}]{2011ApJS..194...28K}
{Knigge}, C., {Baraffe}, I., \& {Patterson}, J. 2011, \apjs, 194, 28

\bibitem[{{Lenz} \& {Breger}(2004)}]{2004IAUS..224..786L}
{Lenz}, P. \& {Breger}, M. 2004, in The A-Star Puzzle, ed. J.~{Zverko},
  J.~{Ziznovsky}, S.~J. {Adelman}, \& W.~W. {Weiss}, Vol. 224, 786--790

\bibitem[{{Mason} {et~al.}(2001){Mason}, {Breeveld}, {Much}, {Carter},
  {Cordova}, {Cropper}, {Fordham}, {Huckle}, {Ho}, {Kawakami}, {Kennea},
  {Kennedy}, {Mittaz}, {Pandel}, {Priedhorsky}, {Sasseen}, {Shirey}, {Smith},
  \& {Vreux}}]{2001A&A...365L..36M}
{Mason}, K.~O., {Breeveld}, A., {Much}, R., {et~al.} 2001, A\&A, 365, L36

\bibitem[{{McDermott} \& {Taam}(1989)}]{1989ApJ...342.1019M}
{McDermott}, P.~N. \& {Taam}, R.~E. 1989, \apj, 342, 1019

\bibitem[{{Mukai}(2017)}]{2017PASP..129f2001M}
{Mukai}, K. 2017, PASP, 129, 062001

\bibitem[{{Nauenberg}(1972)}]{1972ApJ...175..417N}
{Nauenberg}, M. 1972, ApJ, 175, 417

\bibitem[{{Oh} {et~al.}(2018){Oh}, {Koss}, {Markwardt}, {Schawinski},
  {Baumgartner}, {Barthelmy}, {Cenko}, {Gehrels}, {Mushotzky}, {Petulante},
  {Ricci}, {Lien}, \& {Trakhtenbrot}}]{2018ApJS..235....4O}
{Oh}, K., {Koss}, M., {Markwardt}, C.~B., {et~al.} 2018, \apjs, 235, 4

\bibitem[{{Pala} {et~al.}(2022){Pala}, {G{\"a}nsicke}, {Belloni}, {Parsons},
  {Marsh}, {Schreiber}, {Breedt}, {Knigge}, {Sion}, {Szkody}, {Townsley},
  {Bildsten}, {Boyd}, {Cook}, {De Martino}, {Godon}, {Kafka}, {Kouprianov},
  {Long}, {Monard}, {Myers}, {Nelson}, {Nogami}, {Oksanen}, {Pickard},
  {Poyner}, {Reichart}, {Rodriguez Perez}, {Shears}, {Stubbings}, \&
  {Toloza}}]{2022MNRAS.510.6110P}
{Pala}, A.~F., {G{\"a}nsicke}, B.~T., {Belloni}, D., {et~al.} 2022, \mnras,
  510, 6110

\bibitem[{{Parisi} {et~al.}(2012){Parisi}, {Masetti}, {Jim{\'e}nez-Bail{\'o}n},
  {Chavushyan}, {Palazzi}, {Landi}, {Malizia}, {Bassani}, {Bazzano}, {Bird},
  {Charles}, {Galaz}, {Mason}, {McBride}, {Minniti}, {Morelli}, {Schiavone}, \&
  {Ubertini}}]{2012A&A...545A.101P}
{Parisi}, P., {Masetti}, N., {Jim{\'e}nez-Bail{\'o}n}, E., {et~al.} 2012, \aap,
  545, A101

\bibitem[{{Parker} {et~al.}(2005){Parker}, {Norton}, \&
  {Mukai}}]{2005A&A...439..213P}
{Parker}, T.~L., {Norton}, A.~J., \& {Mukai}, K. 2005, \aap, 439, 213

\bibitem[{{Patterson}(1994)}]{1994PASP..106..209P}
{Patterson}, J. 1994, PASP, 106, 209

\bibitem[{{Podsiadlowski} {et~al.}(2003){Podsiadlowski}, {Han}, \&
  {Rappaport}}]{2003MNRAS.340.1214P}
{Podsiadlowski}, P., {Han}, Z., \& {Rappaport}, S. 2003, \mnras, 340, 1214

\bibitem[{{Ramsay}(2000)}]{2000MNRAS.314..403R}
{Ramsay}, G. 2000, MNRAS, 314, 403

\bibitem[{{Rosen} {et~al.}(1988){Rosen}, {Mason}, \&
  {Cordova}}]{1988MNRAS.231..549R}
{Rosen}, S.~R., {Mason}, K.~O., \& {Cordova}, F.~A. 1988, MNRAS, 231, 549

\bibitem[{{Schenker} {et~al.}(2002){Schenker}, {King}, {Kolb}, {Wynn}, \&
  {Zhang}}]{2002MNRAS.337.1105S}
{Schenker}, K., {King}, A.~R., {Kolb}, U., {Wynn}, G.~A., \& {Zhang}, Z. 2002,
  \mnras, 337, 1105

\bibitem[{{Schenker} {et~al.}(2004){Schenker}, {Wynn}, \&
  {Speith}}]{2004ASPC..315....8S}
{Schenker}, K., {Wynn}, G.~A., \& {Speith}, R. 2004, in Astronomical Society of
  the Pacific Conference Series, Vol. 315, IAU Colloq. 190: Magnetic
  Cataclysmic Variables, ed. S.~{Vrielmann} \& M.~{Cropper}, 8

\bibitem[{{Shappee} {et~al.}(2014){Shappee}, {Prieto}, {Grupe}, {Kochanek},
  {Stanek}, {De Rosa}, {Mathur}, {Zu}, {Peterson}, {Pogge}, {Komossa}, {Im},
  {Jencson}, {Holoien}, {Basu}, {Beacom}, {Szczygie{\l}}, {Brimacombe},
  {Adams}, {Campillay}, {Choi}, {Contreras}, {Dietrich}, {Dubberley},
  {Elphick}, {Foale}, {Giustini}, {Gonzalez}, {Hawkins}, {Howell}, {Hsiao},
  {Koss}, {Leighly}, {Morrell}, {Mudd}, {Mullins}, {Nugent}, {Parrent},
  {Phillips}, {Pojmanski}, {Rosing}, {Ross}, {Sand}, {Terndrup}, {Valenti},
  {Walker}, \& {Yoon}}]{2014ApJ...788...48S}
{Shappee}, B.~J., {Prieto}, J.~L., {Grupe}, D., {et~al.} 2014, ApJ, 788, 48

\bibitem[{{Shaw} {et~al.}(2020){Shaw}, {Heinke}, {Mukai}, {Tomsick},
  {Doroshenko}, {Suleimanov}, {Buisson}, {Gandhi}, {Grefenstette}, {Hare},
  {Jiang}, {Ludlam}, {Rana}, \& {Sivakoff}}]{2020MNRAS.498.3457S}
{Shaw}, A.~W., {Heinke}, C.~O., {Mukai}, K., {et~al.} 2020, \mnras, 498, 3457

\bibitem[{{Str{\"u}der} {et~al.}(2001){Str{\"u}der}, {Briel}, {Dennerl},
  {Hartmann}, {Kendziorra}, {Meidinger}, {Pfeffermann}, {Reppin}, {Aschenbach},
  {Bornemann}, {Br{\"a}uninger}, {Burkert}, {Elender}, {Freyberg}, {Haberl},
  {Hartner}, {Heuschmann}, {Hippmann}, {Kastelic}, {Kemmer}, {Kettenring},
  {Kink}, {Krause}, {M{\"u}ller}, {Oppitz}, {Pietsch}, {Popp}, {Predehl},
  {Read}, {Stephan}, {St{\"o}tter}, {Tr{\"u}mper}, {Holl}, {Kemmer}, {Soltau},
  {St{\"o}tter}, {Weber}, {Weichert}, {von Zanthier}, {Carathanassis}, {Lutz},
  {Richter}, {Solc}, {B{\"o}ttcher}, {Kuster}, {Staubert}, {Abbey}, {Holland},
  {Turner}, {Balasini}, {Bignami}, {La Palombara}, {Villa}, {Buttler},
  {Gianini}, {Lain{\'e}}, {Lumb}, \& {Dhez}}]{2001A&A...365L..18S}
{Str{\"u}der}, L., {Briel}, U., {Dennerl}, K., {et~al.} 2001, A\&A, 365, L18

\bibitem[{{Suleimanov} {et~al.}(2016){Suleimanov}, {Doroshenko}, {Ducci},
  {Zhukov}, \& {Werner}}]{2016A&A...591A..35S}
{Suleimanov}, V., {Doroshenko}, V., {Ducci}, L., {Zhukov}, G.~V., \& {Werner},
  K. 2016, \aap, 591, A35

\bibitem[{{Suleimanov} {et~al.}(2019){Suleimanov}, {Doroshenko}, \&
  {Werner}}]{2019MNRAS.482.3622S}
{Suleimanov}, V.~F., {Doroshenko}, V., \& {Werner}, K. 2019, \mnras, 482, 3622

\bibitem[{{Toloza} {et~al.}(2023){Toloza}, {G{\"a}nsicke},
  {Guzm{\'a}n-Rinc{\'o}n}, {Marsh}, {Szkody}, {Schreiber}, {de Martino},
  {Zorotovic}, {El-Badry}, {Koester}, \& {Lagos}}]{2023MNRAS.523..305T}
{Toloza}, O., {G{\"a}nsicke}, B.~T., {Guzm{\'a}n-Rinc{\'o}n}, L.~M., {et~al.}
  2023, \mnras, 523, 305

\bibitem[{{Tomsick} {et~al.}(2023){Tomsick}, {Kumar}, {Coughenour}, {Shaw},
  {Mukai}, {Hare}, {Clavel}, {Krivonos}, {Fornasini}, {Gerber}, \&
  {Joens}}]{2023MNRAS.523.4520T}
{Tomsick}, J.~A., {Kumar}, S.~G., {Coughenour}, B.~M., {et~al.} 2023, \mnras,
  523, 4520

\bibitem[{{Turner} {et~al.}(2001){Turner}, {Abbey}, {Arnaud}, {Balasini},
  {Barbera}, {Belsole}, {Bennie}, {Bernard}, {Bignami}, {Boer}, {Briel},
  {Butler}, {Cara}, {Chabaud}, {Cole}, {Collura}, {Conte}, {Cros}, {Denby},
  {Dhez}, {Di Coco}, {Dowson}, {Ferrando}, {Ghizzardi}, {Gianotti}, {Goodall},
  {Gretton}, {Griffiths}, {Hainaut}, {Hochedez}, {Holland}, {Jourdain},
  {Kendziorra}, {Lagostina}, {Laine}, {La Palombara}, {Lortholary}, {Lumb},
  {Marty}, {Molendi}, {Pigot}, {Poindron}, {Pounds}, {Reeves}, {Reppin},
  {Rothenflug}, {Salvetat}, {Sauvageot}, {Schmitt}, {Sembay}, {Short},
  {Spragg}, {Stephen}, {Str{\"u}der}, {Tiengo}, {Trifoglio}, {Tr{\"u}mper},
  {Vercellone}, {Vigroux}, {Villa}, {Ward}, {Whitehead}, \&
  {Zonca}}]{2001A&A...365L..27T}
{Turner}, M.~J.~L., {Abbey}, A., {Arnaud}, M., {et~al.} 2001, A\&A, 365, L27

\bibitem[{{Vaughan}(2005)}]{2005A&A...431..391V}
{Vaughan}, S. 2005, \aap, 431, 391

\bibitem[{{Warner}(1983)}]{1983ASSL..101..155W}
{Warner}, B. 1983, in Astrophysics and Space Science Library, Vol. 101, IAU
  Colloq. 72: Cataclysmic Variables and Related Objects, ed. M.~{Livio} \&
  G.~{Shaviv}, 155--171

\bibitem[{{Warner}(1995)}]{1995cvs..book.....W}
{Warner}, B. 1995, {Cambridge Astrophysics Series}, Vol.~28, {Cataclysmic
  variable stars} ({Cambridge University Press})

\bibitem[{{Wilms} {et~al.}(2000){Wilms}, {Allen}, \&
  {McCray}}]{2000ApJ...542..914W}
{Wilms}, J., {Allen}, A., \& {McCray}, R. 2000, \apj, 542, 914

\bibitem[{{Woelk} \& {Beuermann}(1996)}]{1996A&A...306..232W}
{Woelk}, U. \& {Beuermann}, K. 1996, \aap, 306, 232

\end{thebibliography}

\begin{appendix}
\section{Spectral parameters obtained using model $A$}
\begin{table}[ht!]
\setlength{\tabcolsep}{4pt}
\renewcommand{\arraystretch}{1.23}
\centering
\caption {Spectral parameters obtained from fitting the average EPIC and RGS spectra of J0826 using model $A$= `\texttt{tbabs$\times$(partcov$\times$tbabs)}$\times$($\sum_{i=1}^4 {\rm \texttt{apec}}_{i}$ + $\sum_{i=1}^6 {\rm \texttt{gauss}}_{i}$ )'.} \label{tab:spec_modelA}
\begin{tabular}{lll}
\hline
Model & Parameters  &  Value \\
\hline
tbabs & N$_{\rm H}$ ($10^{20} \rm ~cm^{-2}$) & $1.6_{-0.6}^{+0.6}$\\
partcov &  cf  & $0.42_{-0.04}^{+0.05}$\\
tbabs & N$_{\rm H,cf}$ ($10^{21} \rm ~cm^{-2}$) & $7.3_{-1.3}^{+1.4} $  \\

apec & T$_{\rm 1}$ (keV)  &  $0.25_{-0.02}^{+0.02} $  \\
& Z$_{1}$ (Z$_{\odot}$) & $2.7_{-0.2}^{+0.2} $  \\
& N$_{\rm 1}$ (10$^{-4}$)   &  $2.4_{-0.4}^{+0.5} $ \\

apec & T$_{\rm 2}$ (keV)  &  $0.90_{-0.03}^{+0.05} $  \\
& Z$_{2}$ (Z$_{\odot}$) & 2.7 (tied to Z$_{1}$)  \\
& N$_{\rm 2}$ (10$^{-4}$)   &  $2.1_{-0.3}^{+0.3} $ \\

apec & T$_{\rm 3}$ (keV)  &  $4.7_{-0.3}^{+0.3} $  \\
& Z$_{3}$ (Z$_{\odot}$) & 2.7 (tied to Z$_{1}$)  \\
& N$_{\rm 3}$ (10$^{-3}$)   &  $3.6_{-0.4}^{+0.4} $ \\

apec & T$_{\rm 4}$ (keV)  &  $46.1_{-4.8}^{+4.7} $  \\
& Z$_{4}$ (Z$_{\odot}$) & 2.7 (tied to Z$_{1}$)  \\
& N$_{\rm 4}$ (10$^{-2}$)   &  $1.61_{-0.03}^{+0.03} $ \\

gauss & $\sigma_{0.43}$ (eV)  &  $1.5_{-0.3}^{+0.4} $  \\
& F$_{\rm 0.43}$ (10$^{-4}$)   &  $2.6_{-0.4}^{+0.5} $ \\
& EW$_{0.43}^{\dagger}$ (eV) &  $17.2_{-4.3}^{+4.0} $\\

gauss & $\sigma_{0.50}$ (eV)  &  $68.8_{-16.9}^{+14.7} $  \\
& F$_{\rm 0.50}$ (10$^{-4}$)   &  $7.1_{-2.2}^{+3.0} $ \\
& EW$_{0.50}^{\dagger}$ (eV) &  $7.8_{-2.2}^{+1.7} $\\

gauss & $\sigma_{0.57}$ (eV)  &  $2.0_{-1.4}^{+2.0} $  \\
& F$_{\rm 0.57}$ (10$^{-4}$)   &  $1.3_{-0.4}^{+0.4} $ \\
& EW$_{0.57}^{\dagger}$ (eV) &  $22.0_{-4.4}^{+5.6} $\\

gauss & $\sigma_{0.65}$ (eV)  &  $7.6_{-3.3}^{+6.2} $  \\
& F$_{\rm 0.65}$ (10$^{-5}$)   &  $4.7_{-3.1}^{+3.2} $ \\
& EW$_{0.65}^{\dagger}$ (eV) &  $22.9_{-3.0}^{+2.6} $\\

gauss & $\sigma_{1.02}$ (eV)  &  $>$ 0.03  \\
& F$_{\rm 1.02}$ (10$^{-5}$)   &  $5.7_{-1.5}^{+1.2} $ \\
& EW$_{1.02}^{\dagger}$ (eV) &  $17.7_{-4.0}^{+3.9} $\\

gauss & F$_{\rm 6.4}$ (10$^{-5}$)   &  $3.4_{-0.3}^{+0.3} $ \\
& EW$_{6.4}$ (eV) &  $86.4_{-0.2}^{+0.2} $\\

\multirow{1}{*}{Unabsorbed X-ray flux} & F$_{\rm X, unabs}$  (10$^{-11}$ erg cm$^{-2}$ s$^{-1}$)    &    $4.32_{-0.01}^{+0.01}$   \\

\multirow{1}{*}{Absorbed X-ray flux} & F$_{\rm X, abs}$  (10$^{-11}$ erg cm$^{-2}$ s$^{-1}$)    &    $3.81_{-0.01}^{+0.01}$   \\

\multirow{1}{*}{Bolometric flux} & F$_{\rm X, bol}$ (10$^{-10}$ erg cm$^{-2}$ s$^{-1}$)    &    $1.087_{-0.005}^{+0.005}$   \\
\multirow{1}{*}{Bolometric luminosity} & L$_{\rm X, bol}$ (10$^{33}$ erg s$^{-1}$)    &    $1.745_{-0.008}^{+0.007}$   \\
& $\chi_\nu^2$ (dof)                       &  1.30 (1234)           \\
\hline
\end{tabular}

\bigskip
\textbf{Note.}  F$_{\rm X, unabs}$ and F$_{\rm X, abs}$ are unabsorbed and absorbed X-ray flux in the 0.3-10.0 keV energy band, respectively; F$_{\rm X, bol}$ is the unabsorbed bolometric flux derived for the 0.001-100.0 keV energy band; L$_{\rm X, bol}$ is the corresponding bolometric luminosity calculated by assuming a distance of 366 pc. 
\newline

$\dagger$ indicates that the EWs are calculated using the model \texttt{`tbabs$\times$(powerlaw+$\sum_{i=1}^5 {\rm \texttt{gauss}}_{i}$})' on the RGS spectrum only.
\end{table}
\end{appendix}
\end{document}